\documentclass[acmsmall]{acmart}
\usepackage[utf8]{inputenc}
\usepackage[T1]{fontenc}
\usepackage[ruled, vlined, linesnumbered]{algorithm2e}
\usepackage{adjustbox}
\usepackage{booktabs}
\usepackage{graphicx}
\usepackage{hyperref}
\usepackage{overpic}
\usepackage{xcolor}
\usepackage{listings}
\usepackage{listings-rust}
\usepackage{microtype}
\usepackage{import}
\usepackage{multicol}
\usepackage{multirow}
\usepackage{pdflscape}
\usepackage{siunitx}
\usepackage{subcaption}
\usepackage{tikz}
\usepackage{xspace}
\usepackage{ifthen}
\usepackage{geometry}

\usepackage[capitalise]{cleveref}

\makeatletter
\newcommand*\Suppressnumber{%
  \lst@AddToHook{OnNewLine}{%
    \let\thelstnumber\relax%
     \advance\c@lstnumber-\@ne\relax%
    }%
}

\SetAlFnt{\ttfamily\footnotesize}

\SetKwIF{If}{ElseIf}{Else}{\texttt{\textbf{if}}}{\texttt{\textbf{then}}}{\texttt{\textbf{else if}}}{\texttt{\textbf{else}}}{\texttt{\textbf{endif}}}

\SetKw{ForEach}{\texttt{\textbf{for each}}}
\SetKwProg{Function}{\texttt{\textbf{function}}}{:}{}
\SetKw{AND}{\texttt{\textbf{and}}}
\SetKw{OR}{\texttt{\textbf{or}}}
\SetKw{NOT}{\texttt{\textbf{not}}}
\SetKw{Continue}{\texttt{\textbf{continue}}}
\SetKw{Return}{\texttt{\textbf{return}}}
\SetKwBlock{DoBlock}{\texttt{\textbf{do}}}{}

\newcommand\boehm{\textsc{BDWGC}\xspace}
\newcommand\rustbacon{\textsc{Bacon-Rajan-CC}\xspace}
\newcommand\rustc{\lstinline{rustc}\xspace}
\newcommand\bronze{\textsc{Bronze}\xspace}
\newcommand\ourgc{\textsc{Alloy}\xspace}
\newcommand\rustgcproj{\textsc{Rust-GC}\xspace}

\newcommand\shifgrethor{\textsc{Shifgrethor}\xspace}

\newcommand\Egcrc{E$_\textrm{GCvs}$\xspace}
\newcommand\Eelision{E$_\textrm{Elision}$\xspace}
\newcommand\Epremopt{E$_\textrm{PremOpt}$\xspace}

\newcommand\rustcversion{1.79.0\xspace}

\lstdefinestyle{rustblock}{
    language=Rust,
    numbersep=5pt,
    xleftmargin=10pt,
    numberstyle=\fontsize{5}{8}\selectfont\tt\color{gray},
    basicstyle=\fontsize{8}{9}\selectfont\tt\color{black},
    keywordstyle=\bfseries, 
    keywordstyle=[2]\color[rgb]{0.75, 0, 0},
    keywordstyle=[3]\color[rgb]{0, 0.5, 0},
    keywordstyle=[4]\color[rgb]{0, 0.5, 0},
    keywordstyle=[5]\color[rgb]{0, 0, 0.75},
    captionpos=b,
    escapeinside={{<!}{!>}},
    numbers=left,
    tabsize=2,
    breakatwhitespace=false,
    breaklines=false,
    showstringspaces=false,
    showspaces=false,
    columns=fullflexible,
    language=Rust
}

\lstdefinelanguage{FsaError}{%
  sensitive%
, morecomment=[l]{//}%
, morecomment=[s]{/*}{*/}%
, morestring=[b]{"}%
, alsodigit={}%
, alsoother={}%
, alsoletter={!}%
, morekeywords={let, fn, self, mut}  
}%

\definecolor{fsared}{rgb}{0.75,0,0}
\definecolor{fsaorange}{rgb}{0.85,0.40,0}

\lstdefinestyle{fsaerror}{
    language=FsaError,
    numbersep=5pt,
    numberstyle=\fontsize{5}{8}\selectfont\tt\color{gray},
    basicstyle=\fontsize{8}{9}\selectfont\tt\color{black},
    keywordstyle=\bfseries, 
    moredelim=**[is][\color{fsared}]{@}{@},
    moredelim=**[is][\color{fsaorange}]{~}{~},
    captionpos=b,
    numbers=left,
    commentstyle=\color{blue!66}\it,
    tabsize=2,
    breakatwhitespace=false,
    breaklines=false,
    showstringspaces=false,
    showspaces=false,
    columns=fullflexible,
    keepspaces=true,
}

\setcopyright{cc}
\setcctype{by}
\acmDOI{10.1145/3763179}
\acmYear{2025}
\acmJournal{PACMPL}
\acmVolume{9}
\acmNumber{OOPSLA2}
\acmArticle{401}
\acmMonth{10}
\received{2025-03-26}
\received[accepted]{2025-08-12}

\begin{document}

\author{Jacob Hughes}
\orcid{0009-0002-5321-0645}
\affiliation{%
  \institution{King's College London}
  \city{London}
  \country{United Kingdom}
}
\email{jh@jakehughes.uk}

\author{Laurence Tratt}
\orcid{0000-0002-5258-3805}
\affiliation{%
  \institution{King's College London}
  \city{London}
  \country{United Kingdom}
}
\email{laurie@tratt.net}

\title{Garbage Collection for Rust: The Finalizer Frontier}
\newcommand\benchmarkcpu{AMD EPYC 7773X 64-Core 3.5GHz CPU\xspace}
\newcommand\benchmarkram{128GiB RAM\xspace}
\newcommand\benchmarkos{Debian 12 (`bookworm')\xspace}
\newcommand\benchmarkpexecs{30\xspace}

\newcommand\somrsast{\textsc{som-rs-ast}\xspace}
\newcommand\somrsbc{\textsc{som-rs-bc}\xspace}
\newcommand\yksom{\textsc{yksom}\xspace}
\newcommand\binarytrees{\textsc{Binary Trees}\xspace}
\newcommand\grmtools{\textsc{grmtools}\xspace}
\newcommand\sws{\textsc{static-web-server}\xspace}
\newcommand\regexredux{\textsc{Regex-Redux}\xspace}
\newcommand\alacritty{\textsc{Alacritty}\xspace}
\newcommand\ripgrep{\textsc{Ripgrep}\xspace}
\newcommand\fd{\textsc{fd}\xspace}

\newcommand\somrsversion{35b780\xspace}
\newcommand\yksomversion{fc7c7c\xspace}
\newcommand\grmtoolsversion{v0.13.4\xspace}
\newcommand\fdversion{v9.0.0\xspace}
\newcommand\ripgrepversion{v14.1.1\xspace}
\newcommand\alacrittyversion{v0.15.0-dev\xspace}
\newcommand\binarytreesversion{Rust\#2\xspace}
\newcommand\regexreduxversion{Rust\#1\xspace}

\newcommand\elisionnaiveworstratio{3.35$\times$\xspace}
\newcommand\gcvsoverallperfratio{1.05$\times$\xspace}
\newcommand\gcvsoverallfootprintratio{2.01$\times$\xspace}
\newcommand\gcvsoverallrssratio{1.34$\times$\xspace}
\newcommand\gcvsgcperfbestratio{0.74$\times$\xspace}
\newcommand\gcvsgcperfworstratio{1.17$\times$\xspace}

\newcommand\alacrittygcownedpct{6.4\%\xspace}
\newcommand\binarytreesgcownedpct{100\%\xspace}
\newcommand\fdgcownedpct{4\%\xspace}
\newcommand\grmtoolsgcownedpct{86\%\xspace}
\newcommand\regexreduxgcownedpct{34\%\xspace}
\newcommand\ripgrepgcownedpct{14.4\%\xspace}
\newcommand\somrsastgcownedpct{42.6\%\xspace}
\newcommand\somrsbcgcownedpct{74\%\xspace}

\newcommand\somrsastboxpct{66\%\xspace}


\newcommand\rc{\lstinline{Rc<T>}\xspace}
\newcommand\arc{\lstinline{Arc<T>}\xspace}
\newcommand\rcboth{\lstinline{Arc<T>}/\lstinline{Rc<T>}\xspace}
\newcommand\gc{\lstinline{Gc<T>}\xspace}
\newcommand\gccfg{\lstinline{Gc<T>} (\textsc{Alloy})\xspace}
\newcommand\gcs{\lstinline{Gc<T>}s\xspace}
\newcommand\rustgc{\lstinline{Gc<T>} (\textsc{Rust-Gc})\xspace}
\newcommand\typedarena{\lstinline{Arena<T>}\xspace}

\begin{abstract}
\noindent Rust is a non-Garbage Collected (GCed) language, but the lack of GC
makes expressing data-structures that require shared ownership awkward,
inefficient, or both. In this paper we explore a new design for, and implementation of, GC in
Rust, called \ourgc. Unlike previous
approaches to GC in Rust, \ourgc allows existing Rust destructors to be automatically
used as GC finalizers: this makes \ourgc integrate better with existing Rust code
than previous solutions but introduces surprising soundness and performance
problems. \ourgc provides novel solutions for the core problems:
\emph{finalizer safety analysis} rejects unsound destructors from
automatically being reused as finalizers; \emph{finalizer elision} optimises
away unnecessary finalizers; and \emph{premature finalizer prevention}
ensures that finalizers are only run when it is provably safe to do so.
\end{abstract}



\maketitle

\section{Introduction}

\begin{figure}[t]
\begin{lstlisting}[
  style=rustblock,
  firstline=6,
  caption={An \ourgc example, showing \lstinline{Gc<T>}
    and destructors as finalizers. We create a type \lstinline{GcNode} which
    models a graph: it stores an 8 bit integer value and a possibly-null
    reference (via Rust's standard \lstinline{Option} type) to a neighbouring node
    (line 1). We add a normal Rust destructor which \ourgc is able to use as a
    finalizer when \lstinline{GcNode} is used inside \lstinline{Gc<T>} (line 2).
    Inside \lstinline{main} we create the first GCed node in the graph (line 5).
    We use Rust's normal \lstinline{RefCell} type to allow the node to be mutated
    (using the \lstinline{RefCell::borrow\_mut} method which dynamically checks
    for mutation that would undermine Rust's static rules)
    and a cycle created directly back to itself (line 6). We then create a second cyclic graph (lines 7
    and 8), immediately assigning it to the \lstinline{gc1} variable (line 9):
    this copies, rather than moves, the \lstinline{Gc<T>}.
    This causes the first cyclic graph \lstinline{GcNode\{value: 1, ..\}}
    to no longer be reachable, so after forcing a collection (line 10) that node
    can be collected. Its finalizer is then scheduled to be run, causing
    \lstinline{drop 1} to be printed out at a later point; when it has completed the GC
    heap memory can be reclaimed. The print statement outputs \lstinline{2 2} (line
    11).},
  label={fig:first_example}
]
struct GcNode { value: u8, nbr: Option<Gc<RefCell<GcNode>>> }
impl Drop for GcNode { fn drop(&mut self) { println!("drop {}", self.value); } }

fn main() {
  let mut gc1 = Gc::new(RefCell::new(GcNode { value: 1, nbr: None } ));
  gc1.borrow_mut().nbr = Some(gc1);
  let gc2 = Gc::new(RefCell::new(GcNode { value: 2, nbr: None } ));
  gc2.borrow_mut().nbr = Some(gc2);
  gc1 = gc2;
  force_gc();
  println!("{} {}", gc1.borrow().value, gc1.borrow().nbr.unwrap().borrow().value);
}
\end{lstlisting}
\end{figure}

Amongst the ways one can classify programming languages are whether they
are Garbage Collected (GCed) or not: GCed languages enable implicit memory management;
non-GCed languages require explicit memory management (e.g~\lstinline{C}'s \lstinline{malloc} /
\lstinline{free} functions). Rust's use of affine types~\citep[p.~5]{pierce04advanced}
and ownership does not fit within this classification: it is not GCed but it has implicit scope-based memory management.
Most portions of Rust programs are as
succinct as a GCed equivalent, but ownership is too inflexible to express
\emph{shared ownership} for data-structures that require multiple owners
(e.g.~doubly linked lists).
Workarounds such as reference counting impose an extra burden on the programmer,
make mistakes more likely, and often come with a performance penalty.

In an attempt to avoid such problems, there are now a number of GCs for Rust
(e.g.~\cite{manish15rustgc, coblenz21bronze, gcarena, boa, shifgrethor}). Most
introduce a user-visible type \lstinline{Gc<T>} which takes a value $v$
of type \lstinline{T} and moves $v$ to the `GC heap'. The \lstinline{Gc<T>}
value itself is a wrapper around a pointer to $v$ on the GC heap.
\lstinline{Gc<T>} can be \emph{cloned} (i.e.~duplicated) and
\emph{dereferenced} to a value of type \lstinline{&T} (i.e.~a type-safe pointer) at will by the user. When no
\lstinline{Gc<T>} wrappers pointing to $v$
can be found, indirectly or directly, from the
program's \emph{roots} (e.g.~variables on the stack),
then the GC heap memory for $v$ can be reclaimed.

It has proven hard to find a satisfying design and implementation for a GC for
Rust, as perhaps suggested by the number of attempts to do so.
We identify two fundamental challenges
for GC for Rust: how to give \lstinline{Gc<T>} an idiomatic and complete
API; and how to make \emph{finalizers} (i.e.~the code that is run just before a
value is collected by the GC) safe, performant, and ergonomic.

In this paper we introduce \ourgc, a new GC for Rust: an example of its use is
shown in \cref{fig:first_example}. \ourgc uses \emph{conservative} garbage
collection (i.e.~treating each reachable machine word as a potential pointer),
which naturally solves the API challenge. However, the finalization challenge
is much more involved: the causes of this challenge, and our solutions to
it, occupy the bulk of this paper.

Normal Rust code uses \emph{destructors} (i.e.~code which is run just before a
value is reclaimed by Rust's implicit memory management) extensively. Although
finalizers and destructors may seem to be synonyms, existing GCs for Rust
cannot reuse destructors as finalizers:
the latter must be manually implemented for each type that needs it.
Unfortunately, even this is trickier than it appears:
it is not possible to implement a finalizer for
\lstinline{Gc<T>} if \lstinline{T} is an external library; some parts of
destructors are automatically created by the Rust compiler, but
hand-written finalizers must duplicate those parts manually; and users
can accidentally cause a type's finalizer to be run more than once. In
short, finalization in existing GCs for Rust is unpalatable.

GCs for Rust are not alone in requiring manually written finalizers.
In a close cousin to our work,
a GC proposal for C++, the reuse of destructors as finalizers was ruled out due to
seemingly insoluble problems~\cite[p.~32]{boehm09garbage}, which we divide
into four categories:
(1) some safe destructors are not safe finalizers;
(2) finalizers can be run prematurely;
(3) running finalizers on the same thread as a paused mutator can cause race conditions and deadlocks;
(4) and finalizers are prohibitively slower than destructors.
All are, at least to some degree, classical GC problems; all are exacerbated
in some way by Rust; and none, with the partial exception of \#2, has
existing solutions.

We show that it is possible to reuse most
Rust destructors as finalizers in a satisfying way. We introduce novel solutions to the
long-standing problems this implies by making use of some of Rust's unusual
static guarantees. We thus gain a better GC for
Rust \emph{and} solutions to open GC problems. Our solutions, in order, are:
(1) \emph{finalizer safety analysis}
extends Rust's static analyses to reject programs whose destructors are not
provably safe to be used as finalizers;
(2) \emph{premature finalizer prevention} automatically inserts fences to prevent
the GC from being `tricked' into collecting values before they are dead;
(3) we run finalizers on a separate thread; and
(4) and \emph{finalizer elision} statically optimises away finalizers if the
underlying destructor duplicates the GC's work.

\ourgc as an implementation
is necessarily tied to Rust, though most of the novel techniques in this paper rely on
general properties of affine types and ownership. While we do not wish
to claim generality without evidence, it seems likely that
many of the techniques in this paper will generalise to other
ownership-based languages, as and when such emerge.

Although \ourgc is not production ready, its performance is already
reasonable: when we control for the (admittedly somewhat slow) conservative GC (\boehm)
\ourgc currently uses, the performance of \ourgc varies from
\gcvsgcperfbestratio to, in the worst case, \gcvsgcperfworstratio that of
reference counting. \ourgc is also sufficiently polished (e.g.~good quality error messages)
in other ways for it to: show a plausible path forwards for those who may
wish to follow it; and to allow others to evaluate whether GC for Rust is a good
idea or not.

This paper is divided into four main parts: GC and Rust background (\cref{sec:background});
\ourgc's basic design (\cref{sec:alloy_design}); destructor and finalizer challenges
and solutions (\cref{sec:destructor challenges,sec:elision,sec:premature_finalize_prevention,sec:fsa}); and evaluation
(\cref{sec:evaluation}). The first three parts have the challenge that our work
straddles two areas that can seem mutually exclusive: GC and Rust. We have
tried to provide sufficient material for readers expert in one of these
areas to gain adequate familiarity with the other, without boring either, but we encourage
readers to skip material they are already comfortable with.

\section{Background}
\label{sec:background}

\subsection{The Challenges of Shared Ownership in Rust}

\begin{figure}[t]
\begin{lstlisting}[
  style=rustblock,
  firstline=5,
  caption={
    A version of~\cref{fig:first_example} using Rust's standard reference
    counting type \lstinline{Rc<T>}. To avoid memory leaks we use \emph{weak}
    references between nodes (line 1). We again create two cyclic graphs (lines
    5--8) using \lstinline{Rc::downgrade} to create weak references (lines 6 and
    8). Since \lstinline{Rc<T>} is not copyable, we must use a manual
    \lstinline{clone} call to have both the \lstinline{rc1} and \lstinline{rc2}
    variables point to the same cyclic graph (line 9). Accessing a neighbour
    node becomes a delicate dance requiring upgrading the weak reference (line 11).
    The need to downgrade \lstinline{Rc<T>} to \lstinline{Weak<T>} and upgrade
    (which may fail, hence the \lstinline{unwrap}) back to \lstinline{Rc<T>}
    creates significant extra complexity relative to~\cref{fig:first_example}: compare
    line 11 in \cref{fig:first_example} to lines 10-11 above.
  },
  label={fig:rc_example}
]
struct RcNode { value: u8, nbr: Option<Weak<RefCell<RcNode>>> }
impl Drop for RcNode { fn drop(&mut self) { println!("drop {}", self.value); } }

fn main() {
  let mut rc1 = Rc::new(RefCell::new(RcNode{value: 1, nbr: None}));
  rc1.borrow_mut().nbr = Some(Rc::downgrade(&rc1));
  let rc2 = Rc::new(RefCell::new(RcNode{value: 2, nbr: None}));
  rc2.borrow_mut().nbr = Some(Rc::downgrade(&rc2));
  rc1 = Rc::clone(&rc2);
  println!("{} {}", rc1.borrow().value,
                      rc1.borrow().nbr.as_ref().unwrap().upgrade().unwrap().borrow().value);
}
\end{lstlisting}
\end{figure}

Rust uses affine types and \emph{ownership}
to statically guarantee that: a value has a
single owner (e.g.~a variable); an owner can \emph{move} (i.e.~permanently
transfer the ownership of) a value to another owner; and
when a value's owner goes out of scope, the value's destructor
is run and its backing memory reclaimed. An owner can pass \emph{references} to a value
to other code, subject to the following static restrictions: there can be
multiple immutable references (`\lstinline{&}') to a value or a single
mutable reference (`\lstinline{&mut}'); and references cannot outlast the owner.
These rules allow many Rust programs to be as succinct as their equivalents
in GCed languages. This suggests that the search for a good GC for Rust may be
intellectually stimulating but of little practical value.

However, there are many programs which need to express data structures
which do not fit into the restrictions of affine types and
ownership. These are often described as `cyclic data-structures', but
in this paper we use the more abstract term `shared ownership', which includes,
but is not limited to, cyclic data-structures.

A common way of expressing shared ownership is to use
the reference counting type \lstinline{Rc<T>} from Rust's
standard library. For many data-structures, this is a reasonable
solution, but some forms of shared ownership require
juggling strong and weak counts. This complicates programs
(see~\cref{fig:rc_example}) and can cause problems when
values live for shorter or longer than intended.

A different solution is to store values in a vector and use
integer indices into that vector. Such indices are morally closer to
machine pointers than normal Rust references: the indices can become
stale, dangle, or may never have been valid in the first place. The programmer
must also manually deal with issues such as detecting unused values,
compaction, and so on. In other words, the programmer
ends up writing a partial GC themselves. A variant of this idea are
\emph{arenas}, which gradually accumulate multiple values but free all of them in one go: values
can no longer be reclaimed too early, though arenas tend to unnecessarily
increase the lifetime of values.

A type-based approach is
\lstinline{GhostCell}~\cite{yanovski21ghostcell}, which uses \emph{branding} to
statically ensure that at any given point only one part of a program can access a
shared ownership data-structure. This necessarily excludes common use cases
where multiple owners (e.g.~in different threads) need to simultaneously access
disjoint parts of a data-structure.

Although it is easily overlooked, some workarounds (e.g.~\lstinline{Rc<T>})
rely on using \emph{unsafe} Rust (i.e.~parts of the language, often involving
pointers, that are not fully statically checked by the compiler). Pragmatically,
we assume that widely used code, even if technically unsafe, has been pored
over sufficiently that it is trustworthy in practise. However,
`new' solutions that a programmer implements using
unsafe Rust are unlikely to immediately reach the same level of trustworthiness.

While we do not believe that every Rust program would be improved by GC, the
variety of workarounds already present in Rust code, and the difficultly
of creating new ones, suggests that there is a subset that would benefit from GC.

\subsection{GC Terminology}

GC is a venerable field and has accumulated terminology that can seem
unfamiliar or unintuitive. We mostly use the same terminology
as~\citet{jones23garbage}, the major parts of which we define here.

A program which uses GC is split between the \emph{mutator} (the user's program) and
the \emph{collector} (the GC itself). At any given point in time, a thread is either
running as a mutator or a collector. In our context, all threads
run as a collector at least sometimes (for reasons that will become apparent
later, some threads always run as a collector).
Tracing and reclamation is performed during a \emph{collection} phase. Our
collections always \emph{stop-the-world}, where all threads running
mutator code are paused while collection occurs.

A \emph{tracing} GC is one that scans memory looking
for reachable values from a program's roots: values , including cycles of
values, that are not reachable from the roots can then be \emph{reclaimed}. In
contrast, a pure reference counting GC does not scan memory, and thus cannot
free values that form a cycle. Increasingly, GC implementations make use of
multiple techniques (see~\cite{bacon04unified}) but, for simplicity's sake,
we assume that implementations wholly use one
technique or another except otherwise stated. For brevity, we use `GC' as a short-hand for `tracing
GC'; when we deal with other kinds of GC (e.g.~reference counting), we
explicitly name them.

We refer to memory which is allocated via \lstinline{Gc<T>} as being on
the \emph{GC heap}. We use the term `GC value' to refer both to the pointer wrapped in a
\lstinline{Gc<T>} and the underlying value on the GC heap, even though multiple
pointers / wrappers can refer to a single value on the GC heap, unless doing so
would lead to ambiguity.

We use `\ourgc' to refer to the combination of: our extension to the Rust
language; our modifications to the \lstinline{rustc} compiler; and our
integration of the Boehm-Demers-Weiser GC (\boehm) into the runtime of programs
compiled with our modified \lstinline{rustc}.

\section{\ourgc: Design and Implementation}
\label{sec:alloy_design}

In this section we outline \ourgc's basic design and implementation choices --
the rest of the paper then goes into detail on the more advanced aspects.

\subsection{Basic Design}
\label{sec:basic design}

\ourgc provides a \lstinline{Gc<T>} type that exposes an API modelled on the
\lstinline{Rc<T>} type from Rust's standard library, because
\lstinline{Rc<T>}: is conceptually
similar to \lstinline{Gc<T>}; widely used in Rust code, and its API
familiar; and that API reflects long-term experience about what Rust programmers
need.

When a user calls \lstinline{Gc::new(v)}, the value \lstinline{v} is
moved to the GC heap: the \lstinline{Gc<T>} value returned to the user is a
simple wrapper around a pointer to \lstinline{v}'s new address. The same underlying GCed value
may thus have multiple, partly or wholly overlapping, references active at any point.
To avoid undermining
Rust's ownership system, dereferencing a \lstinline{Gc<T>}
produces an immutable (i.e.~`\lstinline{&}') reference to the underlying value.
If the user wishes to mutate the underlying value, they must use other Rust
types that enable \emph{interior mutability} (e.g.~\lstinline{RefCell<T>} or
\lstinline{Mutex<T>}).

One feature that \ourgc explicitly supports is the ability in Rust to
cast references to raw pointers and back again. This can occur in two main
ways. \lstinline{Gc<T>} can be dereferenced to \lstinline{&T} which
can then, as with any other reference, be converted to \texttt{*const T}
(i.e.~a C-esque pointer to T). \lstinline{Gc<T>} also supports the common Rust
functions (\lstinline{into_raw} and \lstinline{from_raw}) which wrap
the value-to-pointer conversion in a slightly higher-level API. The ability to
convert references to raw pointers is used in many places (e.g.~Rust's standard
C Foreign Function Interface (FFI)).
We believe that a viable GC for Rust must allow the same conversions,
but doing so has a profound impact because Rust allows raw pointers to be
converted to the integer type \lstinline{usize} and back\footnote{Although
it is outside the scope of this paper, it would
be preferable for Rust to have different types for `data width' and
`address'. Modern C, for example, captures this difference with the \lstinline{size_t} and
\lstinline{uintptr_t} types respectively. Rust now has a provenance lint to
nudge users in this general direction, but the \lstinline{as}
keyword still allows arbitrary conversions.}.

\label{conservative_gc}
Having acknowledged that pointers can be `disguised' as integers, it is then
inevitable that \ourgc must be a conservative GC: if a machine word's integer
value, when considered as a pointer, falls within a GCed block of memory,
then that block itself is considered reachable (and is transitively scanned).
Since a conservative GC cannot know if a word is really a pointer, or is a random sequence of
bits that happens to be the same as a valid pointer, this over-approximates the
\emph{live set} (i.e.~the blocks that the GC will not reclaim). Typically
the false detection rate is very low (see e.g.~a Java study which measures
it at under 0.01\% of the live set~\cite{shahriyar14fast}).

Conservative GC occupies a grey zone in programming language semantics: in most
languages, and most compiler's internal semantics, conservative GC is, formally
speaking, unsound; and furthermore some languages (including Rust) allow
arbitrary `bit fiddling' on pointers, temporarily
obscuring the address they are referring to. Despite this, conservative GC is widely used,
including in the two most widespread web browsers: Chrome uses it in its Blink
rendering engine~\cite{ager13oilpan} and Safari uses it in its JavaScript VM
JavaScriptCore~\cite{pizlo17riptide}. Even in 2025, we lack good alternatives
to conservative GC: there is no cross-platform API for precise GC; and while
some compilers such as LLVM provide some support for GC
features~\cite{llvm14statepoints}, we have found them incomplete and buggy.
Despite the potential soundness worries, conservative GC thus remains a widely
used technique.

\label{gc_is_copyable}
Conservative GC enables \ourgc to make a useful ergonomic improvement over
most other GCs for Rust whose \lstinline{Gc<T>} is only \emph{cloneable}. Such types can be duplicated, but doing
so requires executing arbitrary user code. To make the possible run-time cost of this clear, Rust has
no direct syntax for cloning: users must explicitly call \lstinline{Rc::clone(&v)}
to duplicate a value \lstinline{v}. In contrast, since \ourgc's \lstinline{Gc<T>} is just a wrapper around a pointer it
is not just cloneable but also \emph{copyable}: duplication only requires copying
bytes (i.e.~no arbitrary user code need be executed). Copying is implied by assignment
(i.e. \lstinline{w = v}),
reducing the need for explicit cloning\footnote{The lengthier
syntax \lstinline{y = Gc::clone(&v)} is available, since every copyable type is
also cloneable.}. This is not just a syntactic convenience but also reflects an underlying
semantic difference: duplicating a \lstinline{Gc<T>} in \ourgc is is a cheaper and simpler operation
than most other GCs for Rust which which tend to rely, at least in part, on reference counting.

There is one notable limitation of \lstinline{Gc<T>}'s API relative to
\lstinline{Rc<T>}. The latter, by definition, knows how many references
there are to the underlying data, allowing the value stored inside it
to be mutably borrowed at run-time if there is only a single reference to it
(via \lstinline{get_mut} and \lstinline{make_mut}).
In contrast, \lstinline{Gc<T>} cannot know how many references
there are to the underlying data. As we shall see in~\cref{sec:evaluation}, some Rust programs
are built around the performance advantages of this API (e.g.~turning
`copy on write' into just `write' in some important cases).

\subsection{Basic Implementation}

The most visible aspect of \ourgc is its fork, and extension of, the standard
Rust compiler \rustc. We forked \rustc~\rustcversion, adding
or changing approximately 5,500 Lines of Code (LoC) in the core compiler,
and adding approximately 2,250 LoC of tests.

\ourgc uses \boehm~\cite{boehm88garbage} as the underlying conservative GC, because it is the
most widely ported conservative GC we know of. We use \boehm's \lstinline{GC_set_finalize_on_demand(1)} API,
which causes finalizers to be run on their own thread.

We had to make some minor changes to \boehm to suit our situation.
First, we disabled \boehm's parallel collector
because it worsens \ourgc's performance. It is unclear to us why this happens:
we observe significant lock contention within \boehm during GC
collections, but have not correlated this with a cause.
Second, \boehm cannot scan pointers stored in thread locals
because these are platform dependent. Fortunately, \rustc uses LLVM's
thread local storage implementation, which stores such pointers in the
\lstinline{PT_TLS} segment of the ELF binary: we modified \boehm to scan
this ELF segment during each collection. Third,
\boehm dynamically intercepts thread creation calls so that it can
can scan their stacks, but (for bootstrapping
reasons) is unable to do so in our context: we explicitly changed \ourgc
to register new threads with \boehm.

\section{Destructors and Finalizers}
\label{sec:destructor challenges}

In many GCed languages, `destructor' and `finalizer' are used as synonyms, as
both terms refer to code run when a value's lifetime has ended. In existing GCs
for Rust, these two terms refer to completely different hierarchies of code (i.e.~destructors
and finalizers are fundamentally different). In \ourgc, in contrast, a reasonable first
approximation is that finalizers are a strict subset of destructors. In this section we pick apart
these differences, before describing the challenges of using destructors as
finalizers.

When a value in Rust is \emph{dropped} (i.e.~at the point its owner goes out of lexical
scope) its destructor is automatically run. Rust's destructors enable a style
of programming that originated in C++ called RAII (Resource Acquisition Is
Initialization)~\cite[Section~14.4]{stroustrup97c++}: when a value is dropped,
the underlying resources it possesses (e.g.~file handles or heap memory) are
released. Destructors are used frequently in Rust code (to give a rough idea: approximately 15\%
of source-level types in our benchmark suite have destructors).

Rust destructors are formed of two
parts, run in the following order: a user-defined \emph{drop method}; and
automatically inserted \emph{drop glue}. Drop methods are optional and users
can provide one for a type by implementing the \lstinline{Drop} trait's \lstinline{drop}
method. Drop glue recursively calls destructors of contained types (e.g.~fields
in a \texttt{struct}). Although it is common usage to conflate `destructor' in
Rust with drop methods, drop glue is an integral part of a Rust destructor:
we therefore use `destructor' as the umbrella term for both drop methods and drop glue.

When considering finalizers for a GC for Rust, there are several layers of
design choices. We will shortly see that finalizers cause a number of
challenges (\cref{sec:general_challenges}) and one choice would be to forbid
finalizers entirely. However, this would mean that one could not sensibly embed types
that have destructors in a \lstinline{Gc<T>}. While
Rust does not always call destructors, the situations where this occurs
are best considered `exceptional': not calling destructors from
\lstinline{Gc<T>} would completely undermine reasonable programmer
expectations. Because of this, \ourgc, and indeed virtually all GCs for Rust,
support finalizers in some form.

However, existing GCs force distinct notions of destructors and finalizers onto the programmer.
Where the former have the \lstinline{Drop} trait, the latter typically have
a \lstinline{Finalize} trait. If a user type needs to be finalized then
the user must provide an implementation of the \lstinline{Finalize} trait.
However, doing so introduces a number of problems: (1) external libraries are
unlikely to provide finalizers, so they must be manually implemented
by each consumer; (2) Rust's \emph{orphan
rule}~\cite[Section~6.12]{rustlangref} prevents one implementing traits for
types defined in external libraries (i.e.~unless a library's types were
designed to support \lstinline{Gc<T>}, those types cannot be directly GCed);
(3) one cannot automatically replicate drop glue for finalizers; and (4) one
cannot replicate \rustc's refusal to allow calls to the equivalent of
\lstinline{Drop::drop}.

Programmers can work around problems \#1 and \#2 in various ways. For example,
they can wrap external library types in \emph{newtypes} (zero-cost wrappers)
and implement finalizers on those instead~\cite[Section~19.3]{klabnik18rust}.
Doing so is tedious but not conceptually difficult.

Problem \#3 has partial solutions: for example, ~\cite{manish15rustgc} uses the
\lstinline{Trace} macro to generate \emph{finalizer glue} (the finalizer equivalent of drop glue) for
\texttt{struct} fields. This runs into an unsolvable variant of problem \#2:
types in external libraries will not implement this trait and cannot be
recursively scanned for finalizer glue.

Problem \#4 is impossible to solve in Rust as-is. One cannot define a function
that can never be called --- what use would such a function have? A possible
partial solution might seem to be for the \lstinline{finalize} method take ownership of the value,
but \lstinline{Drop::drop} does not do so because that would not allow drop
glue to be run afterwards.

\subsection{General Challenges When Using Destructors as Finalizers}
\label{sec:general_challenges}

We have stated as our aim that \ourgc should use destructors as finalizers.
Above we explained some Rust-specific challenges --- but there are several
non-Rust-specific challenges too! Fundamentally, finalizers and destructors
have different, and sometimes incompatible, properties. The best
guide to these differences, and the resulting problems, is~\citet{boehm03destructors},
supplemented by later work on support
for GC in the C++ specification~\cite{boehm09garbage}\footnote{These features
were added to the C+11 specification but removed in C++23.}.

An obvious difference between destructors and finalizers is when both
are run. While C++ and Rust define
precisely when a destructor will be run\footnote{Mostly. Rust's `temporary
lifetime extension' delays destruction, but for how long is currently
unspecified.}, finalizers run at an unspecified point in time. This typically
happens at some point after the equivalent destructor would run, though
a program may exit before any given finalizer is run\footnote{A program could pause after `exit' while all queued finalizers are run. We are
not aware of a system which does this.}. There are, however, two situations which
invert this.
First, if a thread exits due to an error, and the program is either not compiled with
unwinding, or the thread has crossed a non-unwinding ABI boundary, then
destructors might not be run at all, where a GC will naturally run the
equivalent finalizers: we do not dwell on this, as both behaviours
are reasonable in their different contexts. Second, and more surprisingly, it is possible for finalizers in
non-error situations to run \emph{prematurely}, that is before the equivalent
destructor~\cite[section~3.4]{boehm03destructors}.

A less obvious difference relates to where destructors and finalizers are run.
Destructors run in the same thread as the last owner of a value.
However, running finalizers in the same thread as the last owner of the value
can lead to race conditions~\cite{niko13destructors} and
deadlocks~\cite[section~3.3]{boehm03destructors} if a finalizer tries to access
a resource that the mutator expects to have exclusive access too.
When such problems affect destructors in normal Rust code, it is the clear result of programmer error, since they should
have taken into account the predictable execution point of destructors. However, since
finalizers do not have a predictable execution point, there is no way
to safely access shared resources if they are run on the same thread.
The only way to avoid this is to run
finalizers on a non-mutator thread --- but not all Rust types / destructors
are safe to run on another thread.

There are several additional differences such as: finalizers
can reference other GCed values that are partly, or wholly, `finalized' and may
have had their backing memory reused; and finalizers can \emph{resurrect} values by
copying the reference passed to the finalizer and storing it somewhere.

Over time, finalizers have thus come to be viewed with increasing suspicion. Java,
for example, has deprecated, and intends eventually removing, per-type
finalizers: instead it has introduced deliberately less flexible per-object `cleaners', whose API
prevents problems such as object resurrection and per-class finalization~\cite{goetz21deprecated}.

\subsection{The Challenge of Finalizers for \ourgc}

At this point we hope to have convinced the reader that: a
viable GC for Rust needs to be able to use existing destructors as finalizers
whenever possible; but that finalizers, even in existing GCs, cause
various problems.

It is our belief that some problems with finalizers are fundamental. For
example, finalizers inevitably introduce latency between the last
use of a value and its finalization.

Some problems with finalizers are best considered the accidental artefacts of
older designs. Java's cleaners, for example, can be thought of as a more
restrictive version of finalizers that allow most common use-cases but forbid
by design many dangerous use cases. For example, per-class/struct finalization
can easily be replaced by per-object/value finalization; and object
resurrection can be prevented if object access requires a level of indirection.
\ourgc benefits from our better shared understanding of such problems and the
potential solutions: it trivially addresses per-object/value finalization
(\lstinline{Gc::new_unfinalizable} function turns finalization off for specific
values) and, as we shall see later, via only slightly more involved means,
object resurrection.

However, that leaves many problems that are potentially in the middle: they are
not obviously fundamental, but there are not obvious fixes for them either. In
our context, where we wish to use destructors as finalizers, four problems have
hitherto been thought insoluble~\cite[p.~32]{boehm09garbage}:
(1) finalizers are prohibitively slower than destructors;
(2) finalizers can run prematurely;
(3) running finalizers on the same thread as a paused mutator can cause race conditions and deadlocks;
(4) some safe destructors are not safe finalizers.

Fortunately for us, Rust's unusual static guarantees, suitably expanded by
\ourgc, allow us to address each problem in novel, satisfying, ways. In the following
section, we tackle these problems in the order above, noting that we tackle problems
\#1 and \#2 separately, and \#3 and \#4 together.

\section{Finalizer Elision}
\label{sec:elision}

As we shall see in \cref{sec:evaluation}, there is a correlation between the
number of finalizers that are run and overhead from GC (with a worst case,
albeit a definite outlier, in
our experiment of \elisionnaiveworstratio slowdown). In this section
we show how to reduce the number of finalizers that are run, which helps
reduce this overhead.

A variety of factors contribute to the finalizer performance overhead, including:
a queue of finalizers
must be maintained, whereas destructors can be run immediately; finalizers run
some time after the last access of a value, making cache misses more likely; and
finalizers can cause values (including values they own) to live for longer
(e.g.~leading to increased memory usage and marking overhead).
Most of these factors are inherent to any GC and
our experience of using and working on \boehm -- a mature, widely used GC -- does
not suggest that it is missing optimisations which would overcome all of this overhead.

Instead, whenever possible, \ourgc \emph{elides} finalizers so that they do not need to be run at all.
We are able to do this because: (1) \boehm is responsible for all allocations and
will, if necessary GC allocations even if they are not directly wrapped in a \lstinline{Gc<T>}; and (2) many Rust destructors
only free memory which \boehm would, albeit with some latency, do anyway.

Consider the standard Rust type \lstinline{Box<T>} which heap allocates space for a value;
when a \lstinline{Box<T>} value is dropped, the heap allocation will be freed.
We can then make two observations. First,
\lstinline{Box<T>}'s drop method solely consists of a \lstinline{deallocate}
call. Second, while we informally say
that \lstinline{Box<T>} allocates on the `heap' and \lstinline{Gc<T>} allocates
on the `GC heap', all allocations in \ourgc are made through \boehm and stored in the same
heap.

When used as a finalizer,
\lstinline{Box<T>}'s drop method is thus unneeded, as the underlying memory will
be freed by \boehm anyway.
This means that there is no need to run a finalizer for a type such as
\lstinline{Gc<Box<u8>>} at all, and the finalizer can be statically elided. However,
we cannot elide a finalizer for a type such as
\lstinline{Gc<Box<Rc<u8>>} because
\lstinline{Rc<T>}'s drop method must be run for the reference count to be decremented.
As this shows, we must consider the complete destructor, and not just the top-level
drop method, when deciding whether a corresponding finalizer can be elided.

\begin{algorithm}[t]
\small
\caption{Finalizer Elision}
\label{alg:elision}
\SetAlgoNoLine
    \Function{\texttt{NeedsFinalizer(T)}}{
        \uIf{\texttt{Impls(T, Drop) \AND \NOT Impls(T, DropMethodFinalizerElidable)}}{
            \Return{\texttt{true}}\;
  }
  \SetAlgoVlined
    \ForEach{field $\in$ T} {
    \DoBlock{
      \SetAlgoNoLine
      \uIf{\texttt{NeedsFinalizer(field)}}{
          \Return{\texttt{true}}\;
      \SetAlgoVlined
     }
    }
  }
    \Return{\texttt{false}}\;
}
\end{algorithm}

\begin{figure}
  \begin{lstlisting}[
    style=rustblock,
    numbers=none,
    label={listing:elision_in_rustc},
    caption={
      A simplified view of how finalizers are elided inside \ourgc. The new compiler intrinsic
      \lstinline{needs_finalizer} returns true if a finalizer is required for a
      type. The \lstinline{Gc<T>} type uses this intrinsic to ensure that the
      value is registered as requiring a finalizer. With optimisations turned on, this
      seemingly dynamic, branching code will be turned into static, branchless code.
    }]
impl<T> Gc<T> {
  pub fn new(value: T) -> Self {
    if needs_finalizer::<T>() { Gc<T>::new_with_finalizer(value) }
    else { Gc<T>::new_without_finalizer(value) }
    ...
  }
}
\end{lstlisting}
\end{figure}

\subsection{Implementing Finalizer Elision}

\label{needs_finalizer_intrinsic}
Finalizer elision statically determines which type's destructors do
not require corresponding finalizers and elides them. It does so conservatively,
and deals correctly with drop glue.

\label{dropmethodfinalizerelidable}
As shown in \cref{alg:elision}, any type which implements
the \lstinline{Drop} trait requires finalization unless it also implements the
new \lstinline{DropMethodFinalizerElidable} \emph{marker trait} (i.e.~a
trait without methods). This trait can be
used by a programmer to signify that a type's drop method need
not be called if the type is placed inside a \lstinline{Gc<T>}. The `Drop'
part of the trait name is deliberate (i.e.~it is not a simplification of
`destructor') as it allows the programmer to reason about a type locally
(i.e.~without considering drop glue or concrete type paramaters).
If the type has a transitively reachable field
whose type implements the \lstinline{Drop} trait but not the
\lstinline{DropMethodFinalizerElidable} trait, then then the top-level type
still requires finalization.

Even though neither normal Rust destructors or \ourgc finalizers are guaranteed
to run, a program whose destructors or finalizers never run would probably not
be usable (leaking resources such as memory, deadlocking, and so on). We
therefore make \lstinline{DropMethodFinalizerElidable} an unsafe
trait, because implementing it inappropriately is likely to lead to undesired
-- though not incorrect! -- behaviour at run-time.

\ourgc modifies the standard Rust library to implement
\lstinline{DropMethodFinalizerElidable} on the following types: \lstinline{Box<T>},
\lstinline{Vec<T>}, \lstinline{RawVec<T>}, \lstinline{VecDeque<T>},
\lstinline{LinkedList<T>}, \lstinline{BTreeMap<K, V>}, \lstinline{BTreeSet<T>},
\lstinline{HashMap<K, V>}, \lstinline{HashSet<T>},
\lstinline{RawTable<K, V>}\footnote{This is a white lie, though the visible effect
is the same. \lstinline{RawTable} is contained in the separate \lstinline{hashbrown} crate
which is then included in Rust's standard library. We previously maintained a
fork of this, but synchronising it is painful. For now, at least, we have hacked explicit knowledge of \lstinline{RawTable}
into the \lstinline{needs_finalize} function.}, and \lstinline{BinaryHeap<T>}. Fortunately,
not only are these types' drop methods compatible with \lstinline{DropMethodFinalizerElidable},
but they are extensively used in real Rust code: they enable significant numbers of
finalizers to be elided.

\cref{listing:elision_in_rustc} shows the new \texttt{const} compiler intrinsic
\lstinline{needs_finalizer} we added to implement \cref{alg:elision}. The
intrinsic is evaluated at compile-time: its result can be inlined into \lstinline{Gc::new},
allowing the associated conditional to be removed too. In other words --
compiler optimisations allowing -- the `does this specific type require a
finalizer?' check has no run-time overhead.

\section{Premature Finalizer Prevention}
\label{sec:premature_finalize_prevention}

Most of us assume that finalizers are always run later than the
equivalent destructor would have run, but they can sometimes run
before~\cite[section~3.4]{boehm03destructors}, undermining soundness.
Such premature finalization is also possible in \ourgc as described thus far
(see~\cref{fig:premature_finalization}). In this section we
show how to prevent premature finalization.

There are two aspects to premature finalization. First, language
specifications often do not define, or do not precisely define, when the earliest point that a value can
be finalized is. While this means that, formally, there is no `premature' finalization,
it seems unlikely that language designers anticipated some of the resulting
implementation surprises (see e.g.~this example in
Java~\cite{shipilev20local}). Second, compiler optimisations -- at least in LLVM --
are `GC unaware', so optimisations such as scalar
replacement can change the point in a program when GCed values appear to be
finalizable.

\begin{figure}[t!]
\begin{lstlisting}[
    style=rustblock,
  caption={An example of possible premature
    finalization. We create a new struct \lstinline{S}
    (line 1) with a drop method that sets the wrapped integer to zero (line 2). In the
    main method, we move an instance of the struct into a
    \lstinline{Box<T>}, which we then move into a \lstinline{Gc<T>} (line 4). We
    obtain a Rust reference to the inner integer (line 5), which at
    run-time will be a pointer to the \lstinline{Box<T>}. At this
    point, the compiler can determine that the \lstinline{Gc<T>} is no longer
    used and overwrite \lstinline{root}'s pointer (which may be in a register). If
    a collection then occurs, a finalizer can run \lstinline{S}'s drop method,
    causing the program to print `0' instead of the expected `1' (line 7).},
    label={fig:premature_finalization}]
struct S { value: u8 }
impl Drop for S { fn drop(&mut self) { self.value = 0; } }

fn main()  {
  let root = Gc::new(Box::new(S{ value: 1 }));
  let inner: &u8 = &**root.value;
  force_gc();
  println!("{}", *inner);
}
\end{lstlisting}
\end{figure}

In our context, it is natural to define premature finalization as a (dynamic) finalizer
for a \lstinline{Gc<T>} value running before the (static) \lstinline{Gc<T>} owner
has gone out of scope. Similar to the high-level proposal mooted
in~\cite[Solution~1]{boehm07optimization}, we must ensure that the dynamic
lifetime of a reference derived from a \lstinline{Gc<T>} matches or
exceeds the lifetime of the \lstinline{Gc<T>} itself.

Our solution relies on adjusting \lstinline{Gc<T>}'s drop method to keep
alive a GCed value for at least the static lifetime of the \lstinline{Gc<T>} itself. In
other words, we ensure that the conservative GC will always see a pointer
to a GCed value while the corresponding \lstinline{Gc<T>} is in-scope.

However, there is a major problem to overcome: copyable types such as
\lstinline{Gc<T>} are forbidden from having destructors. The fundamental
challenge we have to solve is that each copied value will have a destructor
called on it, which has the potential for any shared underlying value to be destructed
more than once. \ourgc explicitly allows \lstinline{Gc<T>} -- but no other
copyable type -- to have a destructor, but to ensure it doesn't cause surprises
in the face of arbitrary numbers of copies, the destructor must be idempotent.
Our task is made easier because \lstinline{Gc<T>} naturally has no drop glue from
Rust's perspective: \lstinline{Gc<T>} consists of a field with a pointer type,
and such types are opaque from a destruction perspective.

We therefore only need to make sure that \lstinline{Gc<T>}'s drop method
is idempotent. Fortunately, this is sufficient for our purposes: we want the drop
method to inhibit finalization but that does not require run-time side effects.
To achieve this, we use a \emph{fence}. These come in various flavours. What
we need is a fence that prevents both: the compiler from reordering
computations around a particular syntactic point; and the CPU from reordering
computations around a particular address. We copy
the platform specific code from the \boehm \lstinline{GC_reachable_here}
macro\footnote{Initially we wrapped this macro in a function.
This led to surprising performance results: an analysis of LLVM IR suggests --
but the sheer quantity of data makes it hard to confirm -- that this may be
due to the effect on inlining heuristics.} into \lstinline{Gc<T>}'s
drop method, which achieves the effect we require.

\subsection{Optimising Premature Finalizer Prevention}
\label{sec:optimising_prem}

The drop method we add to \lstinline{Gc<T>} fully prevents premature
finalization. It also naturally solves a performance problem with the suggested solution
for C++ in~\cite[Solution~1]{boehm07optimization}, which requires keeping alive
all pointers, no matter their type, for their full scope. By definition, our
solution only keeps alive \lstinline{Gc<T>} values: the compiler is free to
optimise values of other types as it so wishes. However, on an
artificial microbenchmark we observed a noticeable overhead from our
fence insertion.

We thus implemented a simple optimisation: we only insert fences for a
\lstinline{Gc<T>} if it has a finalizer. Intuitively, it seems that
we should not generate drop methods in such cases,
but this is difficult to do directly inside \rustc. Instead,
we suppress calls to the drop methods of such types: the two approaches
are functionally equivalent, though ours
does put an extra burden on dead-code elimination in the compiler tool-chain.

\ourgc adds a new pass \lstinline{RemoveElidableDrops} to
\rustc's Mid-Level Intermediate Representation (MIR) processing. MIR is best
thought of as the main IR inside \rustc: it contains the complete set of
functions in the program, where each function consists of a sequence of basic
blocks. Simplifying somewhat, function and drop method calls are represented as
different kinds of \emph{terminators} on basic blocks. Terminators
reference both a callee and a successor basic block.

The \lstinline{remove_elidable_drops} pass iterates over a program's MIR,
identifies drop method terminators which reference elidable finalizers, and turns them into `goto' terminators
to the successor basic basic block. Algorithm 4 in the Appendix presents a
more formal version of this algorithm.

\section{Finalizer Safety Analysis}
\label{sec:fsa}

In this section we address two high-level problems: running finalizers on the same thread
as a paused mutator can cause race conditions and deadlocks; and some safe destructors are not
safe finalizers. Addressing the former problem is conceptually simple --
finalizers must be run on a separate thread -- but we must ensure that doing so
is sound. We therefore consider this a specific instance of the latter problem,
treating both equally in this section.

We therefore introduce Finalizer Safety Analysis (FSA), which prevents
unsafe (in the sense of `not safe Rust') destructors being used as finalizers.
As a first approximation, FSA
guarantees that finalizers are memory safe, cycle safe (i.e.~do not access
already finalized objects), and thread safe. We present the three main
components of FSA individually before bringing them together.

\subsection{Rust References}
\label{sec:fsa_rust_references}

\lstinline{Gc<T>} can store, directly or indirectly, normal Rust
references (i.e.~\lstinline{&} and \lstinline{&mut}), at which point it is
subject to Rust's normal borrow checker rules and cannot outlive the reference.
However, finalizers implicitly extend the lifetime of a GCed value,
including any stored references: accessing a reference in a finalizer could
undermine Rust's borrow checking rules.

A simple way of avoiding this problem is to forbid \lstinline{Gc<T>}
from storing, directly or indirectly, references. This might seem to be
no great loss: storing references in a \lstinline{Gc<T>} largely nullifies
the `GCness' of \lstinline{Gc<T>}. However, we found the result hard to use,
as it can make simple tasks such as gradually migrating existing code
to use \lstinline{Gc<T>} painful.

A moderate, but in our experience insufficient, relaxation is to recognise
that only types that need a finalizer can possibly have problems with
references, and to forbid such types from storing references in
\lstinline{Gc<T>}. For example, if there is no drop method for
\lstinline|struct S{x: &u8}| then its destructor is safe to use as a
finalizer, since its non-drop aspects will not use the \lstinline{&u8} reference.

The eventual rule we alighted upon for FSA is that a destructor for a type
\lstinline{T} can be used as a finalizer provided the destructor's drop methods
do not obtain references derived from \lstinline{T}'s fields (including fields
reachable from its attributes). Using Rust's terminology, we
forbid \emph{projections} (which include a struct's fields, indexes into a
vector, and so on) in destructors from generating references. Any non-projection references that are
used in a destructor are by definition safe to use, as they either exist only
for the duration of the drop method (references to variables on the stack)
or will exist for the remainder of the program (references to global variables).

This rule over-approximates the safe set of destructors. For example, a drop
method that creates a new value and tries to obtain a reference to a field in
it (i.e.~a projection) cannot be a destructor under FSA, even though the
reference cannot outlast the drop method. We found that attempting to relax our
rule further to deal with such cases rapidly complicates exposition and
implementation.

\begin{figure}[ht]
\begin{lstlisting}[
  style=rustblock,
  firstline=5,
  caption={An example of the problems that come from mixing cycles
  and finalization. The salient difference from~\cref{fig:first_example} is that
  the drop method prints the value of a field inside \lstinline{nbr} (line 3). Running this
  program on a strong memory model machine is likely to print either \lstinline{2 0} or \lstinline{1 0}
  depending on whether \lstinline{gc1} or \lstinline{gc2} is finalized first.
  The `seemingly expected' output of \lstinline{1 2} or \lstinline{2 1} would
  never be printed: whichever GCed value is finalized first changes what the
  other GCed value sees in its finalizer. As that implies, this example is
  unsound: whichever finalizer runs second leads to undefined behaviour.},
  label={fig:finalizer_cycle}
]
struct GcNode { value: u8, nbr: Option<Gc<RefCell<GcNode>>> }
impl Drop for GcNode {
  fn drop(&mut self) { self.value = 0; println!("{}", self.nbr.unwrap().borrow().value); }
}

fn main()  {
  let mut gc1 = Gc::new(RefCell::new(GcNode{value: 1, nbr: None}));
  let gc2 = Gc::new(RefCell::new(GcNode{value: 2, nbr: None}));
  gc1.borrow_mut().nbr = Some(gc2);
  gc2.borrow_mut().nbr = Some(gc1);
}
\end{lstlisting}
\end{figure}

\subsection{Cycles and Finalization}
\label{sec:cycles_and_finalization}

One of the main motivations for GCs is that they solve problems with cyclic
data structures. However, finalizers can be unsound if they access
state shared within members of a cycle. \cref{fig:finalizer_cycle} shows an example of undefined
behaviour when two GCed values create a cycle and both their
finalizers reference the other GCed value. Whichever order the finalizers are
run in, at least one of the finalizers will see the other GCed value as partly
or wholly `finalized'.

Most languages and systems we are aware of assume that users either don't run into
this problem (finalization cycles are considered rare in GCed
languages~\cite[p.~229]{jones23garbage}) or know how to deal
with it when they do (e.g.~refactoring the types into parts that do and don't
require finalization~\cite[p.~11]{boehm03destructors}). There is no fully
automatic solution to this problem. Some GCs
offer weak references, which allow users to detect when finalization cycles
have been broken, though they still have to deal with the consequences
manually.

We wanted to provide users with static guarantees that their destructors will
not behave unexpectedly when used as finalizers in a cycle. A first attempt at
enforcing such a property might
seem to be that a \lstinline{Gc<T>} cannot have, directly or indirectly, fields of type
\lstinline{Gc<T>}. This would indeed prevent the mistakes we want to catch but
also disallow shared ownership! We therefore check only that a type's destructor does not,
directly or indirectly, access a \lstinline{Gc<T>}. This allows GCed types to express
shared ownership so long as their destructor(s) do not access other GC types.

To make this check easier to implement, we introduce
an \emph{auto trait}~\cite[Section.~11]{rustlangref},
a kind of marker trait that the compiler propagates automatically.
An auto trait \lstinline{A}
will be automatically implemented for a type \lstinline{T} unless
one of the following is true: there is an explicit \emph{negative
implementation} of \lstinline{A} for \texttt{T}; or \texttt{T}
contains a field that is not itself \lstinline{A}. Informally, we
say that a negative implementation of an auto-trait \emph{pollutes} containing
types.

Our new auto trait is called \lstinline{FinalizerSafe}, and we provide a single
negative implementation \texttt{impl<T>} \lstinline{!FinalizerSafe for Gc<T>}.
This naturally handles transitively reachable code, allowing FSA itself to only
check that a destructor's direct field accesses are \lstinline{FinalizerSafe}.

\subsection{Destructors Need to be Runnable on a Finalizer Thread}
\label{sec:fsa_finalizer_thread}

\begin{figure}[t]
\begin{lstlisting}[
  style=rustblock,
  firstline=10,
  caption={How destructors can cause deadlocks when used as
  finalizers. The mutator creates a reference-counted mutex (line 6),
  placing a copy in a \lstinline{GcNode} that immediately goes out of scope
  (line 7). The mutator then acquires the lock (line 8) but before it
  can release the lock a GC cycle occurs and the \lstinline{GcNode}'s
  finalizer run (line 9). If the finalizer is run on the same thread
  as the mutator, then it will fail to grab the lock (line 2) and cause
  a deadlock.},
  label={fig:finalizer_deadlock}
]
impl Drop for GcNode {
  fn drop(&mut self) { println!("drop {}", self.value.lock().unwrap()); }
}

fn main() {
  let counter = Rc::new(Mutex::new(0));
  { let _ = Gc::new(GcNode { value: Rc::clone(&counter), nbr: None }); }
  let r1 = counter.lock().unwrap();
  force_gc();
  assert_eq!(*r1, 0);
}
\end{lstlisting}
\end{figure}

Running finalizers on the same thread as a mutator can cause problems when the
finalizer accesses state shared with the mutator (see~\cref{sec:general_challenges}
for a general description and \cref{fig:finalizer_deadlock} for a concrete example).
The most general solution to this problem is to run finalizers on a separate
\emph{finalizer thread} that never runs mutator code.

We must therefore ensure that it is safe to run a type's destructor on the
finalizer thread. A conservative definition is that \lstinline{Gc<T>}
is safe to use if \lstinline{T} implements both of Rust's existing \lstinline{Send}
(denoting a type that can be permanently moved from one thread to another)
and \lstinline{Sync} (denoting a type that can be safely accessed simultaneously
by multiple threads) auto traits. However, requiring that finalization be
restricted to types that implement both \lstinline{Send} and
\lstinline{Sync} can be frustrating, particularly because more types
implement \lstinline{Send} than \lstinline{Sync}.

It may seem sufficient for \lstinline{T} to implement \lstinline{Send}
alone so that the value can be safely sent to the finalizer thread. However,
this would not prevent a finalizer indirectly accessing
state shared with a non-GCed value via a mechanism such as \lstinline{Arc},
causing the very problems we are trying to avoid.

FSA thus ignores whether a type
implements \lstinline{Send} or \lstinline{Sync} (or not) and instead examines
the destructor directly. To pass FSA: the destructor must not access thread
locals; and any types the destructor accesses via projections must implement
both \lstinline{Send} and
\lstinline{Sync}. Intuitively, this allows a
non-\lstinline{Send}-or-\lstinline{Sync} type \lstinline{T} to have a safe
finalizer provided that \lstinline{T}'s destructor only access the \lstinline{Send}
and \lstinline{Sync} `subset' of \lstinline{T}.

This rule shows clearly that FSA is a form of abstract interpretation
rather than a mere extension of the type system\footnote{Our initial
implementation of FSA was, in essence, an extension of Rust's type system and
suffered from false positives. For example, porting \alacritty
led to 110 errors --- but our abstract interpretation
approach raises just 2.}. After careful examination we
believe this is compatible with Rust's semantics (and \rustc and LLVM's
implementations) at the time of writing, but it is worth knowing that this rule
would be unsafe in other languages and implementations (for example our
assumption would be unsafe in Java due to synchronisation removal~\cite{wang06escape}). We
leave it as an open question to others as to whether Rust should deliberately permit or forbid
such checks in its semantics.

The implementation of the finalization thread is fairly simple. For example, we
do not need to explicitly synchronise memory between the mutator and
finalization threads because \boehm's stop-the-world collection phase
already synchronises all memory between threads.

\subsection{Putting it All Together}

\begin{algorithm}[t]
\ttfamily
\footnotesize
\caption{Finalizer Safety Analysis}
\label{alg:fsa}

\SetAlgoNoLine
    \Function{\texttt{FinalizerSafetyAnalysis(func)}}{
  \SetAlgoVlined
    \ForEach{\texttt{basic\_block} $\in$ \texttt{func}}{
    \DoBlock{
      t $\gets$ basic\_block.terminator\;
      \SetAlgoNoLine
      \uIf{\NOT \texttt{IsGcConstructorCall(t)}}{
        \Continue\;
      }
      ty $\gets$ GetTyOfGcValue(t)\;
      \uIf{\texttt{isFinalizerUnchecked(ty)} \OR \NOT \texttt{NeedsFinalizer(ty)}}{
        \Continue\;
      }
      \SetAlgoVlined
      \ForEach{\texttt{drop\_method} $\in$ \texttt{GetDropGlue(ty)}}{
        \DoBlock{
          \SetAlgoNoLine
          \uIf{\NOT \texttt{IsMIRAvailable(drop\_method)}}{
              \textit{EmitFinalizerUnsafeError()}\;
          }
          \SetAlgoVlined
          \textit{CheckFunctionSafety(drop\_method)}\;
        }
      }
    }
  }
}

\SetAlgoNoLine
    \Function{\texttt{CheckFunctionSafety(drop)}}{
  \SetAlgoVlined
    \ForEach{\texttt{basic\_block} $\in$ \texttt{drop}}{
    \DoBlock{
        \ForEach{\texttt{statement} $\in$ \texttt{basic\_block}}{
        \DoBlock{
          \ForEach{\texttt{projection} $\in$ \texttt{statement}}{
            \DoBlock{
            \SetAlgoNoLine
              \uIf{\NOT \texttt{IsFinalizerSafe(projection.element)}}{
                  \texttt{EmitFinalizerUnsafeError()}\;
              }
              \SetAlgoVlined
            }
          }
        \SetAlgoNoLine
          \uIf{\texttt{IsFunctionCall(basic\_block.terminator)}}{
              \texttt{CheckFunctionSafety(basic\_block.terminator)}\;
          }
       \SetAlgoVlined
        }
      }
    }
  }
}

\SetAlgoNoLine
\Function{\texttt{IsFinalizerSafe(ty)}}{
    \Return{\texttt{Impls(ty, Send) \AND Impls(ty, Sync) \AND Impls(ty, FinalizerSafe)}}\;
}
\SetAlgoVlined

\end{algorithm}

FSA integrates the seemingly separate components presented above into one. It
iterates over every function in a Rust program analysing destructors of types
that are used in \lstinline{Gc<T>}.
\cref{alg:fsa} shows the essence of FSA
(for example eliding details of caching which \ourgc uses to speed up compile times).

Because FSA is a form of abstract interpretation, we need to determine when to
run FSA on a program. In essence, whenever a previously unchecked type \lstinline{T} is used
to create a new \lstinline{Gc<T>}, FSA is run. As well as the \lstinline{Gc::new}
constructor, \lstinline{Gc<T>} instances can be created with conversion traits
such as \lstinline{From}. We annotated each such entry point with a new \rustc-only
attribute \lstinline{rustc_fsa_entry_point}: calls to functions with this
attribute lead to FSA checks.

A naive implementation of FSA would be a notable cost, so \ourgc uses several
optimisations. As alluded to above, FSA caches the results
of various checks to avoid pointlessly repeating work. We also extend
\lstinline{FinalizerSafe} with negative implementations for \lstinline{&T}, and
\lstinline{&mut T}. If a type
\lstinline{T} implements all of \lstinline{FinalizerSafe}, \lstinline{Send},
and \lstinline{Sync}, we know that there can be no unsafe projections used in a
destructor, and can bypass most FSA checks entirely (though we still need to check
for thread local accesses). Across our benchmark suite, FSA increases compilation
time in release mode by a modest 0.8--1.6\%.

\cref{alg:fsa} also captures \ourgc's approach to error messages.
Rather than just inform a user that `your drop method has not passed FSA', \ourgc
pinpoints which field or line in a drop method caused FSA to fail:
\lstinline{EmitReferenceError} informs the user when a reference in a type
is used in a way that violates FSA (see~\cref{sec:fsa_rust_references}); and
\lstinline{EmitFinalizerUnsafeError} when a drop method
contains code which is unsafe (e.g.~references a \lstinline{Gc<T>} type, an
opaque function, etc.). \cref{fig:finalizer_cycle_err} shows an example of
the errors reported by \ourgc: note that it pinpoints the line within a
drop method that caused an FSA error.

\begin{figure}[t!]
\begin{lstlisting}[
    style=fsaerror,
    xleftmargin=6pt,
    numbers=none,
    caption={The compiler error produced by \ourgc for the example
    in \cref{fig:finalizer_cycle}. This extends \rustc's normal error messages:
    note that both the use of a type with a FSA-incompatible drop method (`has
    a drop method') and the line in the drop method (`caused by the
    expression') are highlighted to the user.},
    label={fig:finalizer_cycle_err}
]
@error: `RefCell::new(GcNode{value: 2, nbr: None})` cannot be safely finalized.@
  --> finalization_cycle.rs:11:21
   |
@7@  |   fn drop(&mut self) { self.value = 0; println!("{}", self.nbr.unwrap().borrow().value); }
   |                                                       @^^^^^^^^@
   |                                              @a finalizer cannot safely dereference this@
                                                  @because it might have already been finalized.@
...
@11@ |   let gc2 = Gc::new(RefCell::new(GcNode{value: 2, nbr: None}));
   |                     @^^^^^^^^^^^^^^^^^^^^^^^^^^^^^^^^^^^^^^^^^@
   			                 @has a drop method which cannot be safely finalized.@
\end{lstlisting}
\end{figure}

\subsubsection{Awkward Kinds of Functions}

FSA can encounter two kinds of `awkward' functions.

First, some functions (e.g.~due to use of trait objects, or FFIs) do not have a
body available when FSA runs: using such a function necessarily causes an FSA
check to fail. One common class of functions which causes this are Rust
intrinsics (e.g.~\lstinline{min_align_of} etc.): we audited the most frequently used of these and
annotated those which are FSA-safe with a new \lstinline{rustc_fsa_safe_fn}
attribute. Other functions whose bodies are unknown cause FSA to fail.

Second, in most cases, FSA runs on Rust functions whose generic types have been
replaced with concrete types (in Rust terminology, functions have been
`monomorphised'). Sometimes, however, FSA encounters functions (e.g.~intrinsics
or functions with certain annotations) whose generic
types have not yet been replaced. FSA can
still run on such functions, but will reject them unless all
generic types imply the \lstinline{FinalizerSafe}, \lstinline{Send}, and
\lstinline{Sync} traits. Note that calling a method on a generically typed
value will lead to FSA finding a method without a body: as in the first case
above, this will cause FSA to fail.

The common theme to both is that we wish FSA to be sound, at which
point we forego completeness. This can cause users frustration when FSA raises
an error on code they know is FSA safe. As is common in Rust, we therefore
provide an unsafe escape hatch which allows users to silence FSA errors when
they can prove to their satisfaction that doing so does undermine correctness.
We experimented with a per-type approach, but found that unduly restrictive: we
therefore provide a per-value escape hatch with the \lstinline{unsafe FinalizerUnchecked<T>}
type. Values wrapped in this type are considered safe to use at all points in
FSA. Our aim is that users should rarely need to resort to this escape hatch,
but, as is not uncommon in Rust, there are valid idioms of use where we found it necessary.

\section{Evaluation}
\label{sec:evaluation}

\begin{table}[t]
\centering
\caption{
    The benchmarks (top) and benchmark suites (bottom) that form our experiment.
    We altered
    them to use different memory allocation strategies (\ourgc, \lstinline{Rc<T>},
    etc.). \binarytrees and \regexredux are classic stand-alone GC benchmarks;
    the other `benchmarks' represent benchmark suites (e.g.~\ripgrep contains
    13 benchmarks). The middle portion of the table shows a variety of `normal' Rust programs;
    the bottom portion of the program shows three implementations of the SOM
    programming language.}
\small
\begin{center}
    
\begin{tabular}{lllll}
\toprule
                  & Version & Description & \#benchmarks \\
\midrule
    \binarytrees & Debian CLBG \binarytreesversion & Heap allocation microbenchmark & 1 \\
    \regexredux & Debian CLBG \regexreduxversion  & Regular expression matching & 1 \\
  \midrule
    \alacritty & \alacrittyversion  & Terminal emulator & 10 \\
    \fd & \fdversion  & Unix find replacement & 7 \\
    \grmtools & \grmtoolsversion  & Lexer / parser library & 4 \\
    \ripgrep & \ripgrepversion  & Fast grep replacement & 13 \\
    \somrsast & git \#\somrsversion  & SOM AST VM & 26 \\
    \somrsbc & git \#\somrsversion  & SOM bytecode VM & 26 \\
\bottomrule
\end{tabular}

\end{center}
\label{tab:benchmarks}
\end{table}

To understand the performance of \ourgc, and the various aspects that make it
up, we carried out three experiments:

\vspace{6pt}
\begin{tabular}{ll}
  \Egcrc & the relative performance of \texttt{Gc<T>} vs.~other allocation strategies \\
  \Eelision & the performance gains of finalizer elision \\
  \Epremopt & the costs of premature finalizer prevention and its optimisation \\
\end{tabular}
\vspace{10pt}

\noindent In this section we explain our methodology and our experimental results.

\subsection{Methodology}

\subsubsection{The Benchmark Suite}

There is no existing benchmark suite for GCs for Rust. Even if such a suite did exist, it
may not have been suitable for our purposes because in experiment \Egcrc we
want to compare programs using existing shared ownership approaches. We searched through roughly the 100 most popular
Rust libraries on \lstinline{crates.io} (the \emph{de facto} standard Rust
package system) looking for suitable candidates. In practise this meant we
looked for crates using reference counting. In the interests of brevity,
for the rest of this section we use `\lstinline{Rc<T>}'
to cover both \lstinline{Rc<T>} and its thread-safe cousin \lstinline{Arc<T>}.

\begin{table}[t]
\centering
\caption{How often relevant types are referenced in source code after our
porting. For \rc, we also show how many weak references are left in:
this is a proxy both for partial porting, and also how the extent
weak references. This is a proxy for the extent of changes that cyclic
data-structures impose upon source code.}
\small
\begin{center}
\begin{tabular}{lrrr}
  \toprule
    & \gc & \rc & \texttt{Weak<T>} \\
  \midrule
  \alacritty   & 107 & 9,450 & 1,970 \\
  \binarytrees &   2 &     0 &     0 \\
  \fd          &   7 &   421 &     1 \\
  \grmtools    & 299 & 1,825 &    23 \\
  \regexredux  & 108 &   109 &     0 \\
  \ripgrep     & 104 &   249 &     4 \\
  \somrsast    & 206 &    35 &     0 \\
  \somrsbc     & 464 &    39 &     0 \\
  \bottomrule
\end{tabular}

        \label{subfig:mem:source}
\end{center}
  \vspace{-8pt}
\end{table}

\cref{tab:benchmarks} shows the resulting suite: note that, except
for \binarytrees and \regexredux, the `benchmarks' are themselves
benchmark suites. Collectively, our suite contains -- depending on whether you
count the SOM implementations' (identical) benchmark suites collectively or
separately -- 62 or 88 benchmarks. \cref{subfig:mem:source} shows how often
relevant types are used after porting. \cref{tab:app:mem:conversion:runtime}
shows the distribution of heap data at run-time. This shows that our suite contains benchmarks
with a variety of memory patterns.

\binarytrees is allocation intensive and sufficiently simple that it
can be easily and meaningfully ported to additional shared ownership strategies:
\rustgcproj, a user library for GC for Rust~\cite{manish15rustgc}; and \typedarena, a non-GC memory
arena~\cite{chiovoloni15typed}.
\alacritty, \fd, and \ripgrep are well known Rust programs, all
of which have their own benchmark suites.
\grmtools is a parsing library which uses \rc extensively
in error recovery: we benchmarked it
using 28KLoC of real Java source code, which we mutated with syntax errors.

SOM is a small, but complete, language in the Smalltalk mould, which has a wide
variety of implementations. Our suite includes two of these: \somrsast
(which represents programs as ASTs); and \somrsbc (which represents programs as
bytecode). Both are existing ports of a Java SOM VM into Rust and use
\rc. We use the same SOM \lstinline{core-lib} benchmarks for
both, derived from git commit \#afd5a6.

We were not able to port all parts of all programs. In particular, some
programs make extensive use of the \lstinline{make_mut} and \lstinline{get_mut}
functions in \lstinline{Rc<T>}, which allow a programmer to mutate their contents
if, at run-time, they only have a single owner. There is not, and cannot be,
equivalent functionality with a copyable \gc type. In some cases we were able
to successfully use alternative mechanisms. In others we judged the usages to
either be rare at run-time (i.e.~not worth porting), or too difficult to port
(i.e.~too much of the program is built around the resulting assumptions). In a
small number of cases we ended up introducing bugs. \alacritty's UTF-8 support
is an example, resulting in deadlocks. Whenever we encountered a bug in our porting,
we reverted back to \lstinline{Rc<T>} for that portion of the port.

\begin{table}[t]
\centering
    \caption{
    Run-time heap distributions. The `Allocated (\#)' columns shows the number
    of values of each type that are allocated (note that most programs
    also allocate values of other types, but those are not shown directly here).
    The `GC Owned' columns shows the proportion of allocated values that
    are owned, directly and indirectly, by \lstinline{Gc<T>} values. 
    For example, a program consisting of a single \lstinline{Gc<Box<T>>} would have
    a `GC Owned' value of 100\% because the \lstinline{Box<T>} is owned by the
    \lstinline{Gc<T>}. As we shall see later, there can be a number of knock-on
    effects when a \lstinline{Gc<T>} owns other such values.}
\small
\begin{center}
\begin{tabular}{lrrrr}
  \toprule
  & \multicolumn{3}{c}{Allocated (\#)} & GC owned (\%) \\
  \cmidrule(lr){2-4}
        & \rc & \texttt{Box<T>} & \gc \\
  \midrule
  \alacritty   & 125      & 8,770        & 2         &  2.70 \\
  \binarytrees & 0        & 3,222,201    & 3,222,190 & 100.00 \\
  \fd          & 17,821   & 306,902      & 61        &   1.23 \\
  \grmtools    & 2,283    & 19,859,431   & 4,038,605 &  44.19 \\
  \regexredux  & 45       & 3,132        & 78        &  15.39 \\
  \ripgrep     & 12,786   & 521,366      & 26,069    &  17.97 \\
  \somrsast    & 15       & 8,586,976    & 1,533,728 &  76.95 \\
  \somrsbc     & 15       & 2,397,931    & 1,530,325 &  99.71 \\
  \bottomrule
\end{tabular}

\label{tab:app:mem:conversion:runtime}
\end{center}
  \vspace{-8pt}
\end{table}

\subsubsection{What We Couldn't Include in the Benchmark Suite}

We tried porting 10 other programs that are not included in our
benchmark suite. To avoid readers wondering if we have `cherry-picked' our
eventual benchmark suite, we briefly report why those other programs have
been excluded. All excluded benchmarks are shown in Table 4 in the Appendix.

Several programs (e.g.~numbat, mini-moka, and salsa), once ported, turned out
to be uninteresting from a GC benchmarking perspective. Irrespective of the
number of source locations that reference memory allocation types, the
benchmarks we could run from them allocated sufficiently little memory that
there are no worthwhile differences between different allocation strategies.
Put another way: these programs are in a sense `the same' from our evaluation perspective.

Two programs (bevy and rust-analyzer) did not run correctly after
porting. Both extensively use the \lstinline{make_mut} or \lstinline{get_mut}
functions in \lstinline{Rc<T>} and reverting those changes made the benchmarks
uninteresting.

\label{rustpython}
We also ported RustPython, but were unable to adjust it to faithfully implement Python-level
destructors. In essence, in RustPython's default configuration, its
representation of objects is not compatible with FSA. This means that we can
not run Python \lstinline{__del__} methods in the finalizer thread. Although
technically this is still compatible with Python's semantics, we felt this
would be a misleading comparison, as our port of RustPython would be doing less
work.

\subsubsection{Running the Experiment}
\label{sec:running_the_experiment}

Our experiment can be seen as a comparison of \ourgc against `normal' Rust.
Fortunately, \ourgc is a strict superset of `normal' Rust: only if users
explicitly opt into GC does \ourgc really become a `GC for Rust'. This allows
us to use the same compiler, standard library, and so on, removing several
potential confounding factor in our results. We compile two binaries:
one without logging features compiled and one with. We only use the latter
when reporting collector related metrics.

A challenge in our experiment is that different allocation strategies can use
different underlying allocators. In particular, \ourgc has to use \boehm, but,
for example, \rc can use a modern allocator such as jemalloc.
Much has changed in the performance of allocators since \boehm's 1980s
roots: in \rc-only benchmarks, we observe an inherent overhead
from \boehm of 2--26\% relative to jemalloc (see Table 6 in the Appendix),
which is a significant, and variable, confounding factor.
Fortunately, \boehm can
be used as a `traditional' allocator that allocates and frees on demand
(i.e.~no conservative GC occurs): in the main experiment, we thus use \boehm as
the allocator for all benchmarks.

We want to understand the memory usage of different allocation strategies over the
lifetime of a benchmark. However, there is no single metric which captures
`memory usage', nor even an agreed set of metrics~\cite{dacapo25}. We use two
metrics to capture different facets: (1) \emph{heap footprint}, the amount of
live heap memory recorded by by Heaptrack~\cite{wolff14heaptrack} at every
allocation and deallocation; and (2) \emph{Resident Set Size} (RSS), the total
physical memory in RAM used by the process (including memory-mapped files,
stack, and code/text segments), sampled at 10Hz. The overhead of recording
heap footprint is much greater than RSS, but it provides a more detailed
view of memory usage.

Another pair of confounding factors are the initial and maximum sizes of the GC heap: too-small values can lead
to frequent resizing and/or `thrashing'; large values to unrealistically few
collections. What `small' and `large' are varies by benchmark,
and `careful' (or thoughtless) choices can significantly distort one's view of
performance. \boehm uses an adaptive strategy by default, growing the heap
size as it detects that it would benefit from doing so. To give some sense of
whether a different strategy and/or heap size would make a difference,
we ran our benchmarks with three different fixed heap sizes. Doing so
either has little effect or speeds benchmarks up; when it does so,
the impact is generally under 10\% and is at most 28\% (the detailed results are presented in
Table 9 in the Appendix). Broadly speaking, this
suggests that \boehm's default heap sizing approach, at least in our context,
is not significantly distorting our view of performance.

We ran each benchmark in our suite 30 times.
We report wall-clock times as returned by the standard Unix \lstinline{time}
utility. The SOM benchmarks are run using its conventional \emph{rebench} tool;
we adjusted \emph{rebench} to use \lstinline{time} for consistency with our
other benchmarks. We ran all benchmarks on an \benchmarkcpu with \benchmarkram,
running \benchmarkos. We turned off turbo boost and hyper-threading,
as both can colour results.

\subsubsection{Data Presentation}
\label{sec:datapres}

Except where otherwise stated, we report means and 99\% confidence intervals
for all metrics. We use the arithmetic mean
for individual benchmarks and the geometric mean for benchmark suites.

When plotting time-series (i.e.~sampled) memory metrics, we face the challenge
that different configurations of the same benchmark can execute at different
speeds. We thus resample each benchmark's
data to 1000 evenly spaced points using linear interpolation. We chose 1000 samples because it is
considerably above the visual resolution of our plots.
After normalization, we calculate the arithmetic mean of the memory footprint
measurement at each grid point (and not the raw underlying data) across all runs of the same benchmark. We record
99\% confidence intervals at each point and show the result as shaded
regions around the mean.

\subsection{Results}

\begin{figure}[t]
    \centering
    \import{figures}{gcvs_perf.pgf}
    \caption{Comparing the effects of \gc and \rc on wall-clock time; heap footprint (i.e.~the size of
    the live set); and RSS. The baseline at 1 is \rc; values less than 1 show
    \gc as better than \rc; and the blue vertical line shows the geometric mean of ratios. The wall-clock times of \gc and
    \rc are similar; the RSS somewhat similar; and the average heap footprint often
    very different. Broadly speaking, \gc increases the average heap footprint
    because GC, and especially finalization, causes values to live for longer.
    Benchmarks which allocate relatively little memory (particularly \ripgrep
    as shown in \cref{tab:app:mem:conversion:runtime}) can
    exaggerate this effect. Perhaps surprisingly, the heap footprint and RSS do not
    correlate. This is partly because the sample rate for RSS is rather low
    (which notably affects fast running benchmarks such as those for \fd) and
    partly because RSS necessarily includes headroom, that is memory beyond that needed
    for the live set (and which may later be returned to the OS).}
    \label{fig:gcvs:summary}
\end{figure}

The main results for \Egcrc can be seen in~\cref{fig:gcvs:summary}. Though
there is variation, \ourgc has an overhead on wall-clock time of 5\% on
our benchmark suite. The
effect on memory is more variable though, unsurprisingly, \ourgc typically has
a larger average heap footprint (i.e.~allocated memory lives for longer). This
metric needs to treated with slight caution: benchmarks which allocate
relatively small amounts of memory (see~\cref{tab:app:mem:conversion:runtime})
can make the relative effect of average heap footprint seem much worse than it
is in absolute terms.

\binarytrees is sufficiently simple that we also used it to compare against
\typedarena and \rustgcproj. The time-series data in~\cref{fig:profiles} is
particularly illuminating (for completeness, Table 5 in the Appendix has the
raw timings). \ourgc is around 3.5$\times$ slower than \typedarena. The
time-series data for the latter shows it going through distinct phases: a (relatively long) allocation
phase, a (relatively moderate) `work' phase, and a (relatively short) deallocation phase.
Put another way: these clear phases make \binarytrees a perfect match for
an arena. In the other approaches, the `work' phase occupies a
much greater proportion of their execution, because it also incorporates
allocator work. \ourgc is around 1.3$\times$ faster than \rc, but both
have similar memory profiles. \ourgc is around 3$\times$ faster than \rustgcproj
and has an average heap footprint around 4$\times$ smaller, reflecting
\ourgc's advantage in not being a user-land library that relies in part on \rc.
Although we caution against over-interpreting a single benchmark, this does give us at least some
idea of the performance ceiling and floor for different approaches.

\begin{figure}[t]
    \import{figures}{profiles.pgf}
    \vspace{-15pt}
    \caption{A selection of time-series data with various GC approaches, showing normalised
    time on the $x$-axis and heap footprint (with 99\% confidence intervals
    shaded) on the $y$-axis.
    (i.e.~the amount of live memory) over time.
    \binarytrees shows an example
    of \ourgc having a comparable heap footprint to \lstinline{Rc<T>};
    \rustgcproj's heap footprint is around 4$\times$ greater. \binarytrees is a
    perfect fit to \typedarena: it frees memory in one batch at the end, and
    because it is 3$\times$ faster than \ourgc, this `wind down' period is
    a substantial portion of the overall (quick!) execution.
    \ripgrep Alternates may seem to be an example of a memory leak in \ourgc,
    but it is really the result of the inevitable delay that GC imposes
    on noticing that values are lived, which is exacerbated by the presence
    of finalizers. The frequent plateaus and dips show that memory is being freed, but at a later
    point than one might initially expect.
    In contrast, \somrsbc JSON Small shows a real memory
    leak due to cyclic objects in \lstinline{Rc<T>}, where \ourgc's heap footprint
    remains steady.}
    \label{fig:profiles}
\end{figure}

The time-series data in~\cref{fig:profiles} helps explain other factors.
For example, it shows that \somrsbc leaks memory on the JSON Small benchmark
(we suspect it also leaks in some other benchmarks, though rarely as visibly).
This is because \rc keeps alive values in cycles; \ourgc does not leak memory on \somrsbc,
as it naturally deals correctly with cycles.

We can see from the time-series data that \ripgrep has a complex heap footprint
pattern. This may suggest a memory leak, but in fact it is a consequence of the
inevitable delay in freeing memory in a GC. In general, GC notices that memory is unused later than reference counting,
but this is exacerbated further by finalizers.
Surprisingly, finalizers can lengthen or shorten an allocation's lifetime. GCed values
with finalizers tend to have longer lifetimes, because they have to wait
in the finalizer queue. However, when a finalizer calls \lstinline{free}
on indirectly owned values, those are immediately marked as not live,
rather than having to wait until the next collection to be discovered as such.
This, albeit indirectly, explains the seemingly random peaks and troughs in
memory usage one can observe in \ripgrep's time-series data.

\begin{figure}[tb]
    \centering
    \import{figures}{elision_perf.pgf}
    \caption{The effects of finalizer elision on various metrics. The top-left
    chart shows the proportion of run-time \gc values that: have had their finalizers elided;
    cannot have their finalizers elided; have no finalizers to elide. This chart
    is best read in conjunction in~\cref{tab:app:mem:conversion:runtime} to (a)
    get a sense of the quantity of run-time memory involved (b) how much indirectly
    owned memory the \gc values have. The other
    plots use `no finalizer elision' as the normalization base (i.e.~values
    below 1 show that finalizer elision improves a metric). Total GC pause
    time is the cumulative time spent in stop-the-world collections. User time
    captures the time spent in all threads, including the finalizer thread.
    Broadly speaking, the more finalizers are elided, and the
    greater the proportion of the overall heap the memory owned by
    \lstinline{Gc<T>}, the better the metrics become.}
    \label{fig:elision:metrics}
\end{figure}

The results of \Eelision are shown in \cref{fig:elision:metrics}. In general,
there is a fairly clear correlation: the more finalizers are removed, and the
greater the proportion of the overall heap the memory owned by
\lstinline{Gc<T>} is, the better the metrics become. However, there are several
caveats. First, when all finalizers are removed, \boehm does not
start a finalizer thread or invoke locking related to it, unduly flattering the time-based
metrics. Second, the quantity of finalizers is only a partial proxy for cost:
some finalizers free up large graphs of indirectly owned values,
which can take some time to run. Third, some benchmarks change the work they do:
\grmtools speeds up so much that its error recovery algorithm has time to do
more work, so while finalizer hugely benefits its GC pause time, its wall-clock
time changes much less. Finally, since finalizers can cause indirectly owned
allocations to be freed earlier than the GC itself does naturally, removing
them can cause indirectly owned values to live for longer: \ripgrep's average heap
footprint highlights this issue.

The results for \Epremopt are shown in \cref{fig:premopt:summary}. We created
three configurations of \ourgc. \emph{None} has no fences, and thus is unsound,
but allows us to approximate (allowing for possible vagaries from running
unsound code!) the best possible outcome. \emph{Naive}
inserts all possible fences. \emph{Optimised} inserts only necessary fences.
Once confidence intervals are taken into account, there are
no statistically significant results for this experiment. Although
it is possible that benchmarking `noise' is hiding a meaningful result, our
data suggests that any such differences are likely to be minimal. To make
up for this disappointment, the fact that there is no difference between
any of these suggests that, on non-artificial benchmarks, premature finalizer
prevention is not a noticeable cost.

\section{Threats to Validity}

We cannot state with certainty that the solutions we have presented are
complete or correct. We have tried to mitigate this by relatively extensive
testing. We created over 100 new tests inside our \rustc fork.
We also ported a number of existing programs to \ourgc --
including programs that we did not expect to make good benchmarks -- since some
GC problems only manifest at scale. However, we could undoubtedly have performed wider
testing, investigated fuzzing, and so on: it is inevitable that there will be
bugs in \ourgc that are entirely our fault. Some of these bugs might have
coloured our performance results.

Our definition of the problems, the design of our solutions, and the way we
implemented them are presented in a fairly informal manner. Though we must
admit that this partly reflects our backgrounds, a truly formal definition
would be challenging for two reasons. First, at the time of
writing, Rust's semantics remain imprecise and incomplete, though this
situation is gradually improving. Second, the problematic
aspects of GC we tackle in this paper have resulted from careful analysis of
running systems (e.g.~\cite{boehm03destructors}): to the best of our knowledge,
no formalism of GC exists which fully covers these aspects. We suspect this is
a mixture of: a lack of research on this problem; the challenging
temporal aspects of GC; and, possibly, a poor fit with existing formalisms.
The closest analogue we know of is the evolving research literature on memory
models which, in a sense, modern GC is a superset of.

\begin{figure}[t]
    \centering
    \import{figures}{premopt_perf.pgf}
    \caption{The effect of premature finalization optimisation,
    normalised to \emph{None}
    (i.e.~no fences). Grey
    bars represent the ratio for \emph{naive} (all possible fences)
    and blue bars \emph{optimised} (obviously unnecessary fences
    removed). Unfortunately, this attempted optimisation has no statistically
    significant effects.}
    \label{fig:premopt:summary}
\end{figure}

Any performance judgements we make are necessarily contingent on our methodology
the benchmark suite we chose, including the proportion of benchmarks that we ported, and
the way we process and present data. For
example, we did not port external libraries to use \lstinline{Gc<T>} so many
benchmarks use a variety of allocation strategies. Even had we ported
everything, we would not be able to say, for example, that finalizer elision will always
improve performance by exactly the factor we see in our experiment: there
undoubtedly exist reasonable, non-pathological, programs which will see
performance changes outside the ranges that our results suggest.

Using \boehm as the allocator for all benchmarks has the advantage of removing
`pure' allocator performance as a confounding factor, but does mean that some
of the performance characteristics of benchmarks will be changed (e.g~due to
the portion of time we spend in the allocator; or \boehm's adaptive heap
sizing strategy). A generic, modern conservative GC, using the insights of
recent non-GC allocators, would almost certainly give different -- though
we suspect not profoundly different -- results.
To the best of our knowledge there is currently no production-quality
modern, generic conservative, GC we could use instead, though we are aware of
at least one attempt to create such an alternative: it will be interesting to rerun our
experiments if and when that arrives.

The RSS memory metric we collect is at Linux's whim: if it does not update as
frequently as we expect, we will see artificially `smoothed' data that may
miss out peaks and troughs. Similar, our interpolation of time-series data onto a normalised
grid can also smooth data. We manually checked a large quantity of data to
ensure this was not a significant effect; by running benchmarks
30 times means it is also less more likely that peaks and troughs are caught at least
sometimes.

\section{Related Work}
\label{sec:related_work}

In this paper we hope to have given sufficient background on GC and the use
of destructors and finalizers in general. In this section we mostly survey the major
parts of the GC for Rust landscape more widely. Our survey is
inevitably incomplete, in part because this is a rapidly evolving field
(a number of changes have occurred since the most recent equivalent
survey we are aware of~\cite{manish21arena}).
We also cover some relevant non-Rust GC work not mentioned elsewhere.

Early versions of Rust had `managed pointers' (using the \lstinline{@T} syntax)
which were intended to represent GC types~\cite{manish21arena}. The core
implementation used reference counting though there were several, sometimes
short-lived, cycle detectors~\cite{hoare22cycles}. Managed pointer support was removed\footnote{In commit
\url{https://github.com/rust-lang/rust/commit/ade807c6dcf6dc4454732c5e914ca06ebb429773}.}
around a year before the first stable release of Rust. This was not the end
of the story for `GC as a core part of Rust', with core Rust developers exploring
the problem space in more detail~\cite{felix15specifying, felix16roots,
manish16gc}. Over time these efforts dwindled, and those interested in GC for Rust largely moved from anticipating
\rustc support to expecting to have to do everything in user-level libraries.

One of the earliest user-level GC for Rust libraries is
\rustbacon~\cite{rustbacon}. This provides a type \lstinline{Cc<T>}
which is similar in intention to \ourgc's \lstinline{Gc<T>}.  The
mechanism by which objects are collected is rather different: they have
a naive reference count, which causes objects outside a cycle to
have deterministic destruction; and users can manually invoke a cycle
detector, which uses trial deletion in the style of \citet{bacon01concurrent}\footnote{While
the term `trial deletion' does not appear in the original paper, it is now widely used
to describe part of the algorithm.} to identify objects in unused cycles. Cycle detection requires
users manually implementing a \texttt{Trace} trait which
traverses a type's fields. Destructors
are used as finalizers: to avoid the problems with Rust references we solved in
\cref{sec:fsa_rust_references}, \rustbacon imposes a \lstinline{T:'static}
lifetime bound on the type parameter passed to \lstinline{Cc<T>}. Simplifying
slightly, this means that any references in such a type must be valid for the
remaining lifetime of the program, a severe restriction. Unlike our approach to the access
of already-finalized values (\cref{sec:cycles_and_finalization}), it can only detect
such accesses at runtime, leading to a (safe) Rust \lstinline{panic}.

\label{rustgc}
Probably the best known GC for Rust is \rustgcproj~\cite{manish15rustgc} (partly
covered in \cref{sec:destructor challenges}). \rustgcproj's \lstinline{Gc<T>} provides
a similar API to \ourgc, with the notable exception that its \lstinline{Gc<T>}
is not, and cannot be, copyable, thus always requiring calls to
\lstinline{Gc::clone}. Although, like \ourgc, \rustgcproj allows \lstinline{Gc<T>} values to
be converted into pointers, its lack of conservative GC means that users
must ensure that a \lstinline{Gc<T>} wrapper is kept alive for the entire
lifetime of pointers derived from it. Similarly to \rustbacon,
GCed values are reference counted, with occasional
tracing sweeps to identify cycles, though \rustgcproj performs cycle detection
automatically (i.e.~it doesn't require manual calls to a function such as
\lstinline{collect_cycles}). Drop methods are not used as finalizers: if a
finalizer is required, a manual implementation of the \lstinline{Finalize} trait
must be provided; finalizer glue can be largely, though not fully (see
\cref{sec:destructor challenges}), automatically created by the provided
\lstinline{Trace} macro. \rustgcproj detects accesses to already-finalized values
dynamically at run-time, panicking if they occur. Unlike \rustbacon, these accesses
are detected by recording what the collector's state is in: if the
collector is in a `sweep' phase, any access of a \lstinline{Gc<T>} leads to a
panic. We have not yet verified whether cross-thread collection / sweeping can
evade this check.

An example of moving beyond reference counting in a GC for Rust is
\shifgrethor~\cite{shifgrethor}. It requires \lstinline{Gc} values to
be created by a \lstinline{Root<'root>}: the resulting \lstinline{Gc<'root, T>}
is then tied to the lifetime of the
\lstinline{Root<'root>}. This allows roots to be precisely identified, but
requires explicitly having access to a \lstinline{Root<'root>} whenever a
\lstinline{Gc<'root, T>} is used. As with \rustgcproj, \shifgrethor requires users
to manually implement a \lstinline{Finalize} trait, though \shifgrethor's is
more restrictive: not only can other GCed values not be accessed (implicitly
solving the same problem as \cref{sec:cycles_and_finalization}) but any other
type without the same \lstinline{'root} lifetime as the GCed value is
forbidden. This means that many seemingly safe finalizers require
implementing the unsafe \lstinline{UnsafeFinalize} trait. We view
\shifgrethor as proof that accurately tracking GC roots
in normal Rust without reference counting is possible, though it cannot
deal with references being converted into pointers and \lstinline{usize}s.

A different means of tackling the root-finding problem is \textsc{GcArena}~\cite{gcarena},
which uses branding in a similar way to
\lstinline{GhostCell}s (see \cref{sec:background}). In essence, users provide
a special `root' type which is the only place where roots can be stored.
Mutating the heap can only be done in the context of functions that are
passed a branded reference to the GCed heap. Once such a function has completed,
\textsc{GcArena} is in full control of the GC heap, and knows that
only the root type needs to be scanned for roots. This leads to a precise
guarantee about GC reference lifetimes. However, if code executes in an
arena for too long, the system can find itself starved of resources,
with no way of recovering, even if much of the arena is no longer used. \textsc{GcArena} was originally
part of the \emph{Piccolo} VM  (which was itself previously called
\emph{Luster}), a Lua VM written in Rust. Such VMs have a frequently executed
main loop which is a natural point for a program to relinquish references
to the GCed heap, but this is not true of many other GCed programs.

One attempt to improve upon \rustgcproj is \bronze~\cite{coblenz21bronze},
though it shows how challenging it can be to meaningfully improve GC for Rust: both of
its main advances have subsequently been disabled because they are not just unsound
but actively lead to crashes. First, \bronze tried to solve the root-finding problem by using LLVM's \lstinline{gc.root} intrinsic
at function entries to generate stack-maps (a run-time mechanism for accurately
tracking active pointers). This rules out the false positives that are
inevitable in conservative GC. However, \bronze
could not track nested references: if a \lstinline{Gc<T>} was used as a field in a struct, it
was not tracked by the GC. Second, \bronze tried to give GC in
Rust similar semantics to non-ownership languages such as Java. It did this by
allowing shared mutation, undermining Rust's borrow checker.

Chrome's rendering engine \emph{Blink} uses the conservative GC
\textsc{Oilpan}. It has the interesting property that it has two classes of
finalizers. `Full finalizers' are similar to finalizers in \ourgc, running on a
finalizer thread at an indeterminate future point, but with the difference that
they can only reference parts of a GCed value. To mitigate this,
`pre-finalizers' are run by the collector on the same thread as mutator
as soon as an object as recognised as unused, and can
access all of a GCed value. Pre-finalizers are necessary, but not encouraged, because they
implicitly pause the stop-the-world phase of the collector.
This reflects the fact that latency is a fundamental concern
for a rendering engine: \ourgc currently makes no pretences to being low latency.

\section{Conclusions}

We introduced a novel design for GC in Rust that solves a number of outstanding
challenges in GC for Rust, as well as -- by taking advantage of Rust's
unusual static guarantees -- some classical GC finalizer problems.
By making integration with existing Rust code easier than previous GCs for
Rust, we hope to have shown a pragmatic route for partial or wholesale
migration of Rust code that would benefit from GC.

Challenges and future opportunities remain. For example, \ourgc
is an `all or nothing' cost: if you want to use \lstinline{Gc<T>} in a single
location, you must pay the costs of the GC runtime and so on. \ourgc's absolute
speed is, we believe, limited by \boehm: it is probable that using a
semi-precise GC and/or a faster conservative GC could change our view of the
absolute performance speed

In summary, we do not claim that \ourgc is the ultimate design
for GC in Rust -- reasonable people may, for example, disagree on whether the costs of
conservative GC are worth the gains -- but it does
show what can be achieved if one is willing to alter the
language's design and \rustc.

\section*{Data Availability Statement}

The accompanying artefact~\cite{hughes25garbageartefact} contains: the source
code necessary to run this paper's experiment (including generating figures
etc.) from scratch; and data from a run of the experiment that we used in this
paper.

\begin{acks}
This work was funded by an EPSRC PhD studentship and the Shopify / Royal
Academy of Engineering Research Chair in Language Engineering. We thank
Steve Klabnik and Andy Wingo for comments.
\end{acks}

\clearpage

\bibliographystyle{ACM-Reference-Format}
\bibliography{bib}

\clearpage

\appendix

\section*{Appendices}

\begin{algorithm}[h]
\small
\caption{Removing elidable drops}
  \SetAlgoNoLine
    \Function{\texttt{RemoveElidableDrops(func)}}{
      \SetAlgoVlined
        \ForEach {\texttt{basic\_block} $\in$ \texttt{func}}{
            \DoBlock{
          \SetAlgoNoLine
            \uIf {\texttt{IsDropTerminator(basic\_block.terminator.kind)}} {
                ty $\gets$ GetTypeOfDroppedValue(block.terminator)\;
                \uIf {\texttt{IsGcType(ty)}}{
                    \uIf {\NOT \texttt{RequiresFinalizer(ty)}} {
                        ReplaceTerminator(basic\_block)\;
                    }
                }
            }
          \SetAlgoVlined
        }
        }
    }
  \SetAlgoNoLine
    \Function{\texttt{ReplaceTerminator(basic\_block)}}{
        drop\_func $\gets$ GetDropFunc(basic\_block.terminator)\;
        last\_block $\gets$ GetLastBasicBlock(drop\_func)\;
        block.terminator $\gets$ last\_block.terminator\;
    }
  \label{alg:barrier_removal}
\end{algorithm}

\clearpage

\section*{Additional Experiment Data}

In this appendix, we provide supplementary plots for our performance evaluation. These are as follows:
\\\\

\textbf{Misc.}
\begin{itemize}
  \item \cref{tab:app:exclusions} (page~\pageref{tab:app:exclusions}): Rust programs excluded from our benchmark suite.
\end{itemize}
\textbf{\Egcrc: a performance evaluation of \ourgc vs Reference Counting.}

\begin{itemize}
  \item \cref{tab:gcvs:raw} (page~\pageref{tab:gcvs:raw}): Wall-clock times for the \Egcrc experiment.
  \item \cref{tab:app:gcvs:alloc:cost} (page~\pageref{tab:app:gcvs:alloc:cost}): A comparison of the allocation overhead of \boehm vs \texttt{jemalloc}.
  \item \cref{tab:app:gcvs:wallclock} (page~\pageref{tab:app:gcvs:wallclock}): Raw wall-clock time data for each benchmark.
  \item \cref{tab:app:gcvs:user} (page~\pageref{tab:app:gcvs:user}): Raw user time data.
  \item \cref{tab:app:gcvs:fixed:heap}
      (page~\pageref{tab:app:gcvs:fixed:heap}): Comparison between \boehm’s
default adaptive heap sizing and fixed size heap configurations.\end{itemize}
\textbf{\Eelision: a performance evaluation of \ourgc's finalizer elision.}

\begin{figure}[htbp]
    \centering
    \begin{subfigure}{1\textwidth}
        \centering
        \includegraphics[width=1\textwidth,keepaspectratio]{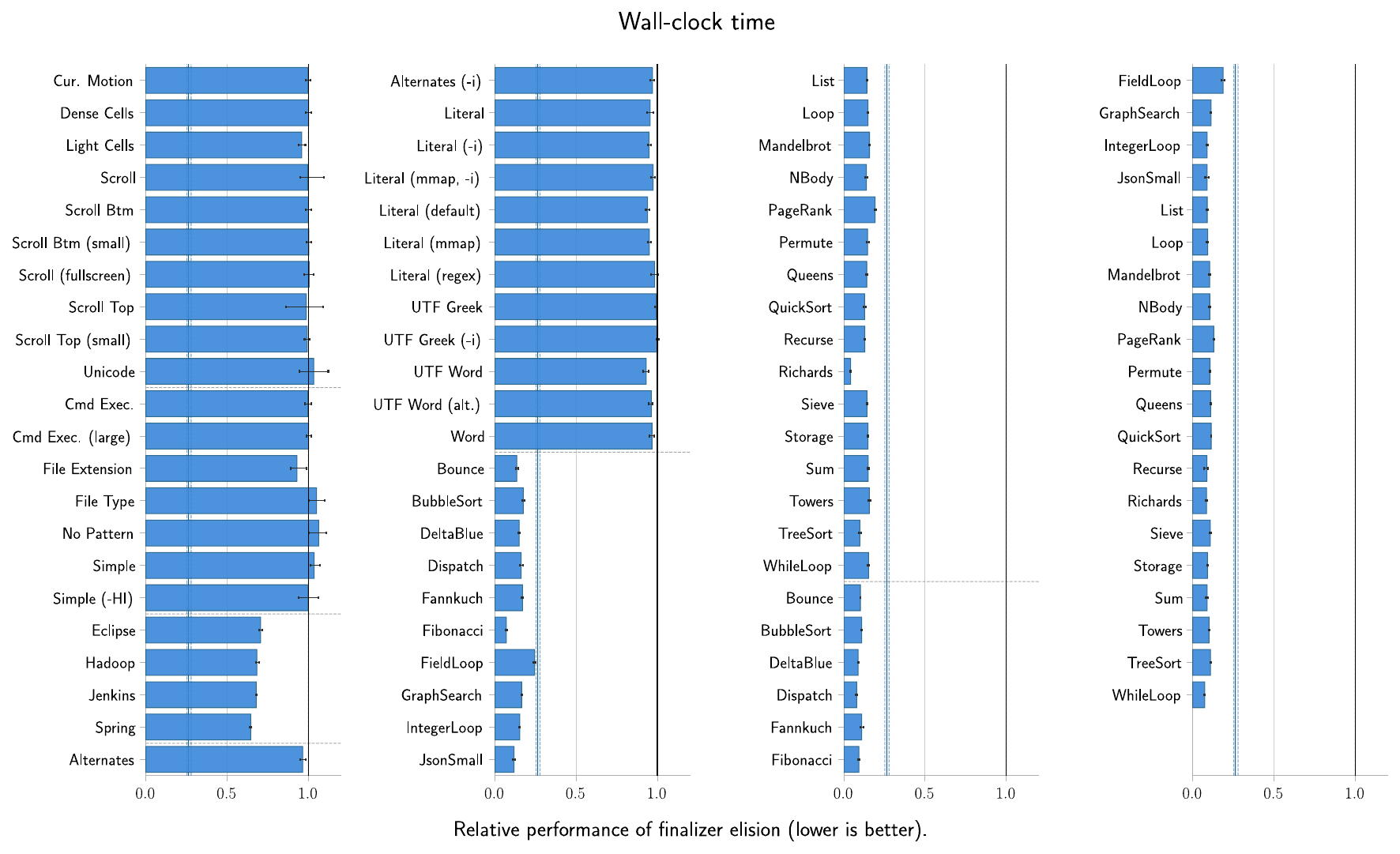}
    \end{subfigure}%
    \hfill
    \begin{subfigure}{1\textwidth}
        \centering
        \includegraphics[width=1\textwidth,keepaspectratio]{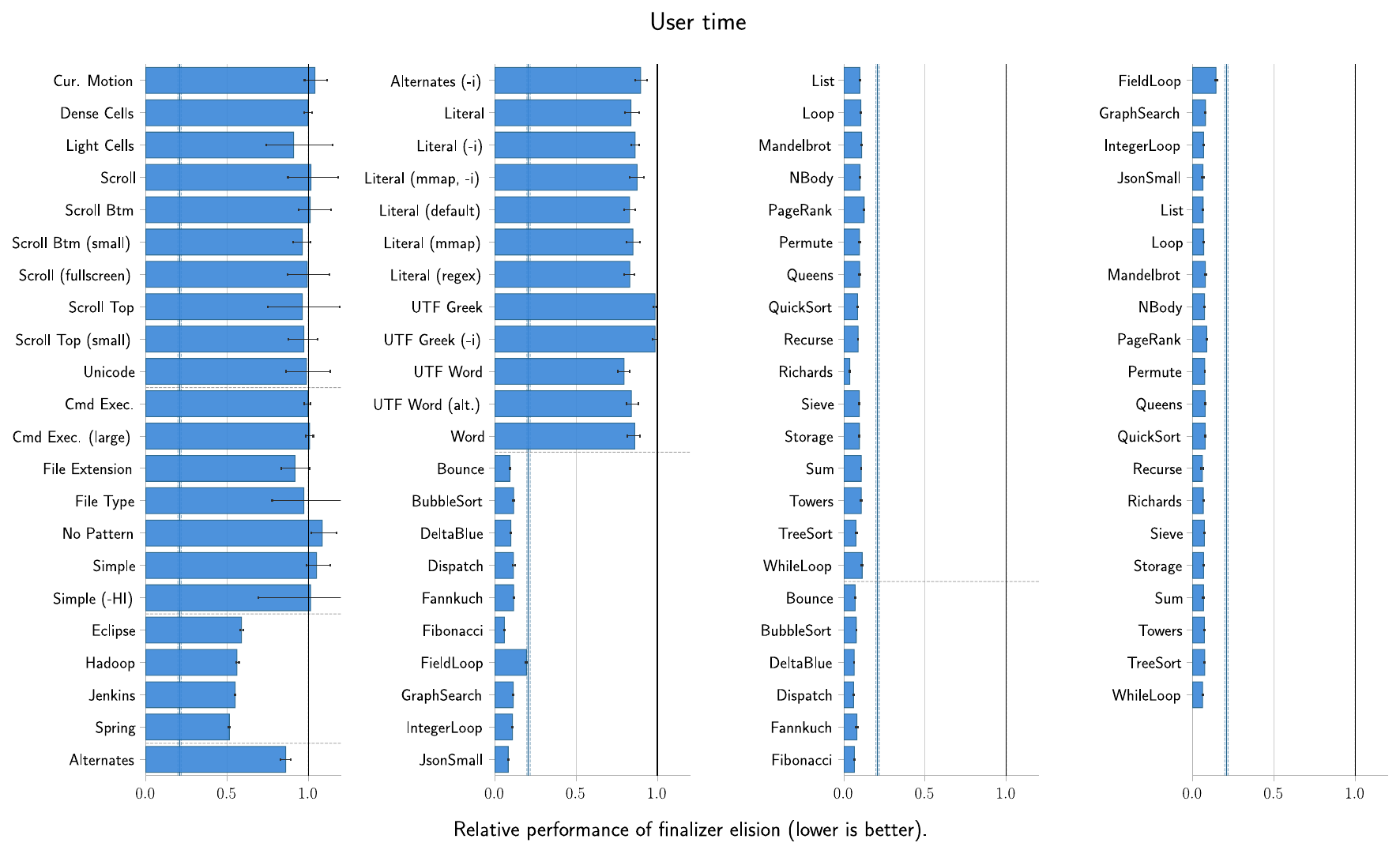}
    \end{subfigure}
    \caption{Wall-clock and user time performance comparison for finalizer
    elision on each benchmark. The bars show the relative performance of \ourgc
    after applying our elision optimization, normalized against the baseline
    (solid black line). The vertical blue line marks the overall geometric mean
    (with shaded area for CIs). User time often shows greater improvement than
    wall-clock time, as elision reduces the CPU overhead of the finalization
    thread.}
    \label{fig:app:elision:perf}
\end{figure}

\begin{itemize}
  \item \cref{tab:app:elision:pct} (page~\pageref{tab:app:elision:pct}): Percentage of finalizers \ourgc elided.
  \item \cref{fig:app:elision:perf} (page~\pageref{fig:app:elision:perf}): Wall-clock and user time comparison.
  \item \cref{tab:app:elision:wallclock} (page~\pageref{tab:app:elision:wallclock}): Raw wall-clock time data.
  \item \cref{tab:app:elision:user} (page~\pageref{tab:app:elision:user}): Raw user time data.
  \item \cref{tab:app:elision:heap} (page~\pageref{tab:app:elision:heap}): Raw average heap usage data.
\end{itemize}

\clearpage

\begin{table}[t]
\centering
\caption{Rust programs excluded from our benchmark suite after attempted porting to \ourgc.}
\small
\begin{center}
\scalebox{0.8}{%
\begin{tabular}{lll}
\toprule
Benchmark         & Description                              & Reason for exclusion                                       \\
\midrule
bevy          & ECS game engine in Rust                   & Unable to port successfully (see \cref{rustpython}) \\
dyon          & Scripting language in Rust                & Unable to port successfully (see \cref{rustpython})\\
jiff          & A datetime library for Rust               & Too few allocations to measure \\
mini-moka     & Concurrent in-memory cache library        & Too few allocations to measure \\
numbat        & Math search engine                        & Too few allocations to measure \\
rkyv          & Zero-copy deserialization framework       & Insufficient \texttt{Gc<T>} coverage in benchmarks \\
RustPython    & Python interpreter written in Rust        & Difficulty retro-fitting \texttt{\_\_del\_\_} semantics (see \cref{rustpython}). \\
rust-analyzer & Language server for Rust                  & Unable to port successfully (see \cref{rustpython}) \\
salsa         & Incremental recomputation library         & Too few allocations to measure \\
WLambda       & Scripting language written in Rust        & Insufficient \texttt{Gc<T>} coverage in benchmarks \\
\bottomrule
\end{tabular}

}
\label{tab:app:exclusions}
\end{center}
\end{table}

\begin{table}[h]
    \centering
    \caption{Wall-clock times (in seconds) for the \Egcrc experiment, comparing
    different memory management strategies across our benchmark suites.}
    \begin{tabular}{lrrrr}
\toprule
& \multicolumn{4}{c}{Wall-clock time (s)} \\
\cmidrule(lr){2-5}
& \gc & \rc & \rustgc & \typedarena \\
\midrule
\alacritty   & 0.41 \scriptsize\textcolor{gray!60}{[0.39,0.45]} & 0.40 \scriptsize\textcolor{gray!60}{[0.38,0.44]} & -- & -- \\
\binarytrees & 0.11 \scriptsize\textcolor{gray!60}{[0.11,0.11]} & 0.15 \scriptsize\textcolor{gray!60}{[0.14,0.15]} & 0.33 \scriptsize\textcolor{gray!60}{[0.32,0.33]} & 0.03 \scriptsize\textcolor{gray!60}{[0.03,0.04]} \\
\fd          & 0.33 \scriptsize\textcolor{gray!60}{[0.29,0.38]} & 0.31 \scriptsize\textcolor{gray!60}{[0.26,0.37]} & -- & -- \\
\grmtools    & 3.06 \scriptsize\textcolor{gray!60}{[3.00,3.14]} & 3.24 \scriptsize\textcolor{gray!60}{[3.17,3.31]} & -- & -- \\
\regexredux & 0.47 \scriptsize\textcolor{gray!60}{[0.47,0.47]} & 0.45 \scriptsize\textcolor{gray!60}{[0.45,0.46]} & -- & -- \\
\ripgrep     & 1.61 \scriptsize\textcolor{gray!60}{[1.55,1.69]} & 1.52 \scriptsize\textcolor{gray!60}{[1.45,1.59]} & -- & -- \\
\somrsast  & 0.92 \scriptsize\textcolor{gray!60}{[0.88,0.95]} & 0.79 \scriptsize\textcolor{gray!60}{[0.76,0.82]} & -- & -- \\
\somrsbc   & 0.28 \scriptsize\textcolor{gray!60}{[0.27,0.29]} & 0.29 \scriptsize\textcolor{gray!60}{[0.28,0.30]} & -- & -- \\
\bottomrule
\end{tabular}

    \label{tab:gcvs:raw}
\end{table}

\begin{table}[h]
    \centering
    \caption{Wall-clock times (in seconds) comparing the jemalloc
    allocator and the \boehm allocator from the \boehm across our
    benchmark suite.}
    \label{tab:app:gcvs:alloc:cost}
\scalebox{0.9}{%
  \begin{tabular}{lrlrlr}
  \toprule
      & \multicolumn{2}{c}{jemalloc} &
        \multicolumn{3}{c}{\boehm} \\
  \cmidrule(lr){2-3}\cmidrule(lr){4-6}
    & \multicolumn{2}{c}{Wall-clock (s)}
    & \multicolumn{2}{c}{Wall-clock (s)}
    & Slowdown \\
\midrule
  \alacritty & 0.39 & \scriptsize\textcolor{gray!60}{[0.36,0.43]} &
               0.40 & \scriptsize\textcolor{gray!60}{[0.38,0.43]} &
               1.02$\times$ \\
    \binarytrees & 0.09 & \scriptsize\textcolor{gray!60}{[0.08,0.10]} & 0.11 & \scriptsize\textcolor{gray!60}{[0.12, 0.08]} & 1.22$\times$ \\
  \fd        & 0.28 & \scriptsize\textcolor{gray!60}{[0.25,0.34]} &
               0.31 & \scriptsize\textcolor{gray!60}{[0.27,0.35]} &
               1.08$\times$ \\
  \grmtools   & 2.89 & \scriptsize\textcolor{gray!60}{[2.82,2.97]} &
               3.10 & \scriptsize\textcolor{gray!60}{[3.01,3.18]} &
               1.07$\times$ \\
    \regexredux & 0.47 & \scriptsize\textcolor{gray!60}{[0.46,0.48]} & 0.49 & \scriptsize\textcolor{gray!60}{[0.48, 0.50]} & 1.04$\times$ \\
  \ripgrep    & 1.43 & \scriptsize\textcolor{gray!60}{[1.36,1.51]} &
               1.46 & \scriptsize\textcolor{gray!60}{[1.39,1.53]} &
               1.02$\times$ \\
  \somrsast   & 0.58 & \scriptsize\textcolor{gray!60}{[0.56,0.60]} &
               0.73 & \scriptsize\textcolor{gray!60}{[0.70,0.75]} &
               1.26$\times$ \\
  \somrsbc    & 0.23 & \scriptsize\textcolor{gray!60}{[0.22,0.24]} &
               0.27 & \scriptsize\textcolor{gray!60}{[0.26,0.29]} &
               1.18$\times$ \\
  \midrule
  Overall     & 0.49 & \scriptsize\textcolor{gray!60}{[0.47,0.51]} &
               0.56 & \scriptsize\textcolor{gray!60}{[0.53,0.58]} &
               1.14$\times$ \\
  \bottomrule
\end{tabular}

}
\end{table}

\begin{table}[t]
    \centering
  \caption{Wall-clock execution times (in seconds) for each benchmark in the
    \Egcrc experiment. Values show arithmetic means over \benchmarkpexecs runs
    and include 99\% confidence intervals. Because \binarytrees and \regexredux
    are stand-alone benchmarks they are omitted here; their timings appear in
    \cref{tab:gcvs:raw}.}
  \begin{subtable}{0.49\textwidth}
    \centering
\scalebox{0.7}{%
  \begin{tabular}{llrlrlrl}
\toprule
& & \multicolumn{6}{c}{Wall-clock time (s)} \\
\cmidrule(lr){3-8}
& & \multicolumn{2}{c}{jemalloc}
& \multicolumn{4}{c}{\boehm} \\
\cmidrule(lr){3-4}\cmidrule(lr){5-8}
& & \multicolumn{2}{c}{\rc} &
\multicolumn{2}{c}{\gc} &
\multicolumn{2}{c}{\rc} \\
\midrule
\multirow{10}{*}{\rotatebox{90}{\alacritty}} & Cur. Motion & 0.65 & \scriptsize\textcolor{gray!60}{$\pm$0.02} & 0.66 & \scriptsize\textcolor{gray!60}{$\pm$0.01} & 0.66 & \scriptsize\textcolor{gray!60}{$\pm$0.01} \\
 & Dense Cells & 2.16 & \scriptsize\textcolor{gray!60}{$\pm$0.03} & 2.16 & \scriptsize\textcolor{gray!60}{$\pm$0.02} & 2.14 & \scriptsize\textcolor{gray!60}{$\pm$0.04} \\
 & Light Cells & 0.37 & \scriptsize\textcolor{gray!60}{$\pm$0.01} & 0.37 & \scriptsize\textcolor{gray!60}{$\pm$0.01} & 0.38 & \scriptsize\textcolor{gray!60}{$\pm$0.01} \\
 & Scroll & 0.25 & \scriptsize\textcolor{gray!60}{$\pm$0.01} & 0.26 & \scriptsize\textcolor{gray!60}{$\pm$0.02} & 0.26 & \scriptsize\textcolor{gray!60}{$\pm$0.01} \\
 & Scroll (fullscreen) & 0.38 & \scriptsize\textcolor{gray!60}{$\pm$0.01} & 0.39 & \scriptsize\textcolor{gray!60}{$\pm$0.01} & 0.38 & \scriptsize\textcolor{gray!60}{$\pm$0.01} \\
\bottomrule
\end{tabular}

}
  \end{subtable}
  \hfill
  \begin{subtable}{0.49\textwidth}
    \centering
\scalebox{0.7}{%
  \begin{tabular}{llrlrlrl}
\toprule
& & \multicolumn{6}{c}{Wall-clock time (s)} \\
\cmidrule(lr){3-8}
& & \multicolumn{2}{c}{jemalloc}
& \multicolumn{4}{c}{\boehm} \\
\cmidrule(lr){3-4}\cmidrule(lr){5-8}
& & \multicolumn{2}{c}{\rc} &
\multicolumn{2}{c}{\gc} &
\multicolumn{2}{c}{\rc} \\
\midrule
\multirow{4}{*}{\rotatebox{90}{\grmtools}} & Eclipse & 3.48 & \scriptsize\textcolor{gray!60}{$\pm$0.03} & 3.44 & \scriptsize\textcolor{gray!60}{$\pm$0.03} & 3.74 & \scriptsize\textcolor{gray!60}{$\pm$0.04} \\
 & Hadoop & 2.86 & \scriptsize\textcolor{gray!60}{$\pm$0.01} & 2.92 & \scriptsize\textcolor{gray!60}{$\pm$0.01} & 3.09 & \scriptsize\textcolor{gray!60}{$\pm$0.01} \\
 & Jenkins & 2.90 & \scriptsize\textcolor{gray!60}{$\pm$0.00} & 2.99 & \scriptsize\textcolor{gray!60}{$\pm$0.05} & 3.02 & \scriptsize\textcolor{gray!60}{$\pm$0.01} \\
 & Spring & 2.43 & \scriptsize\textcolor{gray!60}{$\pm$0.00} & 2.51 & \scriptsize\textcolor{gray!60}{$\pm$0.01} & 2.63 & \scriptsize\textcolor{gray!60}{$\pm$0.02} \\
\midrule
\multirow{13}{*}{\rotatebox{90}{\ripgrep}} & Alternates & 1.18 & \scriptsize\textcolor{gray!60}{$\pm$0.01} & 1.29 & \scriptsize\textcolor{gray!60}{$\pm$0.02} & 1.21 & \scriptsize\textcolor{gray!60}{$\pm$0.01} \\
\bottomrule
\end{tabular}

      }
  \end{subtable}
    \label{tab:app:gcvs:wallclock}
\end{table}

\begin{table}[t]
    \centering
  \caption{User times (in seconds) for each benchmark in the \Egcrc experiment.
    Values show arithmetic means over \benchmarkpexecs runs and include 99\%
    confidence intervals.}
  \begin{subtable}{0.49\textwidth}
    \centering
\scalebox{0.7}{%
  \begin{tabular}{llrlrlrl}
\toprule
& & \multicolumn{6}{c}{User time (s)} \\
\cmidrule(lr){3-8}
& & \multicolumn{2}{c}{jemalloc}
& \multicolumn{4}{c}{\boehm} \\
\cmidrule(lr){3-4}\cmidrule(lr){5-8}
& & \multicolumn{2}{c}{\rc} &
\multicolumn{2}{c}{\gc} &
\multicolumn{2}{c}{\rc} \\
\midrule
\multirow{10}{*}{\rotatebox{90}{\alacritty}} & Cur. Motion & 1.08 & \scriptsize\textcolor{gray!60}{$\pm$0.05} & 1.09 & \scriptsize\textcolor{gray!60}{$\pm$0.04} & 1.09 & \scriptsize\textcolor{gray!60}{$\pm$0.04} \\
 & Dense Cells & 6.66 & \scriptsize\textcolor{gray!60}{$\pm$0.13} & 6.62 & \scriptsize\textcolor{gray!60}{$\pm$0.12} & 6.58 & \scriptsize\textcolor{gray!60}{$\pm$0.15} \\
 & Light Cells & 0.34 & \scriptsize\textcolor{gray!60}{$\pm$0.03} & 0.34 & \scriptsize\textcolor{gray!60}{$\pm$0.03} & 0.34 & \scriptsize\textcolor{gray!60}{$\pm$0.03} \\
 & Scroll & 0.11 & \scriptsize\textcolor{gray!60}{$\pm$0.01} & 0.11 & \scriptsize\textcolor{gray!60}{$\pm$0.01} & 0.12 & \scriptsize\textcolor{gray!60}{$\pm$0.02} \\
 & Scroll (fullscreen) & 0.24 & \scriptsize\textcolor{gray!60}{$\pm$0.02} & 0.28 & \scriptsize\textcolor{gray!60}{$\pm$0.03} & 0.22 & \scriptsize\textcolor{gray!60}{$\pm$0.02} \\
\bottomrule
\end{tabular}

}
  \end{subtable}
  \hfill
  \begin{subtable}{0.49\textwidth}
    \centering
\scalebox{0.7}{%
  \begin{tabular}{llrlrlrl}
\toprule
& & \multicolumn{6}{c}{User time (s)} \\
\cmidrule(lr){3-8}
& & \multicolumn{2}{c}{jemalloc}
& \multicolumn{4}{c}{\boehm} \\
\cmidrule(lr){3-4}\cmidrule(lr){5-8}
& & \multicolumn{2}{c}{\rc} &
\multicolumn{2}{c}{\gc} &
\multicolumn{2}{c}{\rc} \\
\midrule
\multirow{4}{*}{\rotatebox{90}{\grmtools}} & Eclipse & 3.32 & \scriptsize\textcolor{gray!60}{$\pm$0.03} & 3.28 & \scriptsize\textcolor{gray!60}{$\pm$0.03} & 3.59 & \scriptsize\textcolor{gray!60}{$\pm$0.04} \\
 & Hadoop & 2.77 & \scriptsize\textcolor{gray!60}{$\pm$0.01} & 2.80 & \scriptsize\textcolor{gray!60}{$\pm$0.01} & 2.99 & \scriptsize\textcolor{gray!60}{$\pm$0.01} \\
 & Jenkins & 2.74 & \scriptsize\textcolor{gray!60}{$\pm$0.01} & 2.81 & \scriptsize\textcolor{gray!60}{$\pm$0.04} & 2.90 & \scriptsize\textcolor{gray!60}{$\pm$0.01} \\
 & Spring & 2.35 & \scriptsize\textcolor{gray!60}{$\pm$0.01} & 2.40 & \scriptsize\textcolor{gray!60}{$\pm$0.01} & 2.54 & \scriptsize\textcolor{gray!60}{$\pm$0.02} \\
\midrule
\multirow{13}{*}{\rotatebox{90}{\ripgrep}} & Alternates & 0.42 & \scriptsize\textcolor{gray!60}{$\pm$0.02} & 0.51 & \scriptsize\textcolor{gray!60}{$\pm$0.02} & 0.43 & \scriptsize\textcolor{gray!60}{$\pm$0.02} \\
\bottomrule
\end{tabular}

      }
  \end{subtable}
    \label{tab:app:gcvs:user}
\end{table}

\begin{table}[h]
  \caption{The effects of fixing the heap size (see~\cref{sec:running_the_experiment})
    on wall-clock time, reported as ratios relative to \boehm’s default
    adaptive RSS strategy. We chose per-benchmark heap sizes based on our
    observations of their memory usage, though some benchmarks fail at smaller
    values. Broadly speaking, these results show that the performance
    differences due to different heap sizes are relatively minor, and that \boehm's adaptive
    strategy has not unduly coloured our results.}
  \begin{tabular}{lrrl}
    & Heap Size (MiB) & Relative wall-clock time & Benchmarks failed \\
\hline
    \multirow{3}{*}{\alacritty}      & 16  & 0.96 \scriptsize\textcolor{gray!60}{[0.91, 0.99]} &  \\
                                    & 32  & 0.98 \scriptsize\textcolor{gray!60}{[0.95, 1.02]} &  \\
                                    & 64  & 0.94 \scriptsize\textcolor{gray!60}{[0.89, 0.98]} &  \\
\midrule
    \multirow{3}{*}{\binarytrees}   & 4  & 0.88 \scriptsize\textcolor{gray!60}{[0.82, 1.02]} &  \\
                                    & 8  & 0.90 \scriptsize\textcolor{gray!60}{[0.80, 1.01]} &  \\
                                    & 16 & 0.87 \scriptsize\textcolor{gray!60}{[0.82, 0.94]} &  \\
\midrule
    \multirow{3}{*}{\fd}             & 16  & 0.94 \scriptsize\textcolor{gray!60}{[0.90, 0.99]} &  \\
\bottomrule
\end{tabular}

\label{tab:app:gcvs:fixed:heap}
\end{table}

\clearpage

\begin{table}[t]
    \centering
    \caption{Percentage of finalizers \ourgc was able to elide for each benchmark.}
  \begin{subtable}{0.49\textwidth}
    \centering
\scalebox{0.8}{%
  \begin{tabular}{llr}
\toprule
    & & \multicolumn{1}{c}{Fin. elided (\%)} \\
\midrule
\multirow{10}{*}{\rotatebox{90}{\alacritty}}
& Cur. Motion & 0.00  \\
 & Dense Cells & 0.00  \\
 & Light Cells & 0.00  \\
 & Scroll & 0.00  \\
 & Scroll Btm & 0.00  \\
 & Scroll Btm (small) & 0.00  \\
 & Scroll (fullscreen) & 0.00  \\
 & Scroll Top & 0.00  \\
 & Scroll Top (small) & 0.00  \\
 & Unicode & 0.00  \\
\multirow{6}{*}{\rotatebox{90}{\fd}}
 & Cmd Exec. & 8.93  \\
 & Cmd Exec. (large) & 9.09  \\
 & File Extension & 72.73  \\
 & File Type & 24.54  \\
 & No Pattern & 0.04  \\
 & Simple & 61.54  \\
 & Simple (-HI) & 61.54  \\

\midrule
\multirow{26}{*}{\rotatebox{90}{\somrsast}}
 & Bounce & 100.00  \\
 & BubbleSort & 100.00  \\
 & DeltaBlue & 100.00  \\
 & Dispatch & 100.00  \\
 & Fannkuch & 100.00  \\
 & Fibonacci & 100.00  \\
 & FieldLoop & 100.00  \\
 & GraphSearch & 99.99  \\
 & IntegerLoop & 100.00  \\
 & JsonSmall & 100.00  \\
 & List & 100.00  \\
 & Loop & 100.00  \\
 & Mandelbrot & 100.00  \\
 & NBody & 99.99  \\
 & PageRank & 99.99  \\
 & Permute & 100.00  \\
 & Queens & 100.00  \\
 & QuickSort & 100.00  \\
 & Recurse & 100.00  \\
 & Richards & 100.00  \\
 & Sieve & 100.00  \\
 & Storage & 100.00  \\
 & Sum & 100.00  \\
 & Towers & 99.99  \\
 & TreeSort & 99.99  \\
 & WhileLoop & 100.00  \\
\bottomrule
\end{tabular}

}
  \end{subtable}
  \hfill
  \begin{subtable}{0.49\textwidth}
    \centering
\scalebox{0.8}{%
  \begin{tabular}{llr}
\toprule
    & & \multicolumn{1}{c}{Fin. elided (\%)} \\
\midrule
\multirow{4}{*}{\rotatebox{90}{\grmtools}}
 & Eclipse & 11.79  \\
 & Hadoop & 32.99  \\
 & Jenkins & 26.74  \\
 & Spring & 37.25  \\
\midrule
\multirow{13}{*}{\rotatebox{90}{\ripgrep}}
 & Alternates & 79.04  \\
 & Alternates (-i) & 79.04  \\
 & Literal & 79.04  \\
 & Literal (-i) & 79.04  \\
 & Literal (mmap, -i) & 79.04  \\
 & Literal (default) & 79.04  \\
 & Literal (mmap) & 79.04  \\
 & Literal (regex) & 79.04  \\
 & UTF Greek & 79.04  \\
 & UTF Greek (-i) & 79.04  \\
 & UTF Word & 79.04  \\
 & UTF Word (alt.) & 79.04  \\
 & Word & 79.04  \\
\midrule
\multirow{26}{*}{\rotatebox{90}{\somrsbc}}
 & Bounce & 72.50  \\
 & BubbleSort & 76.34  \\
 & DeltaBlue & 73.86  \\
 & Dispatch & 77.78  \\
 & Fannkuch & 72.74  \\
 & Fibonacci & 70.37  \\
 & FieldLoop & 83.33  \\
 & GraphSearch & 76.56  \\
 & IntegerLoop & 75.00  \\
 & JsonSmall & 71.85  \\
 & List & 79.41  \\
 & Loop & 74.82  \\
 & Mandelbrot & 70.93  \\
 & NBody & 86.17  \\
 & PageRank & 70.95  \\
 & Permute & 79.23  \\
 & Queens & 79.04  \\
 & QuickSort & 69.33  \\
 & Recurse & 82.61  \\
 & Richards & 71.17  \\
 & Sieve & 73.15  \\
 & Storage & 73.84  \\
 & Sum & 75.00  \\
 & Towers & 78.29  \\
 & TreeSort & 65.19  \\
 & WhileLoop & 74.98  \\
\bottomrule
\end{tabular}

      }
  \end{subtable}
    \label{tab:app:elision:pct}
\end{table}

\clearpage

\clearpage

\begin{table}[t]
    \centering
    \caption{Wall-clock execution times (seconds) for each benchmark in the
    \Eelision experiment, shown before and after applying \ourgc's
    finalizer elision optimisation. Values show arithmetic means over
    \benchmarkpexecs runs, with 99\% confidence intervals.}
  \begin{subtable}{0.49\textwidth}
    \centering
\scalebox{0.8}{%
\begin{tabular}{llrlrl}
\toprule
& & \multicolumn{4}{c}{Wall-clock time (s)} \\
\cmidrule(lr){3-6}
 &  & \multicolumn{2}{c}{Before elision} & \multicolumn{2}{c}{After elision} \\
\midrule
\multirow{10}{*}{\rotatebox{90}{\alacritty}} & Cur. Motion & 0.66 & \scriptsize\textcolor{gray!60}{$\pm$0.01} & 0.66 & \scriptsize\textcolor{gray!60}{$\pm$0.01} \\
 & Dense Cells & 2.17 & \scriptsize\textcolor{gray!60}{$\pm$0.03} & 2.17 & \scriptsize\textcolor{gray!60}{$\pm$0.02} \\
 & Light Cells & 0.39 & \scriptsize\textcolor{gray!60}{$\pm$0.01} & 0.38 & \scriptsize\textcolor{gray!60}{$\pm$0.01} \\
 & Scroll & 0.27 & \scriptsize\textcolor{gray!60}{$\pm$0.01} & 0.27 & \scriptsize\textcolor{gray!60}{$\pm$0.01} \\
 & Scroll (fullscreen) & 0.39 & \scriptsize\textcolor{gray!60}{$\pm$0.01} & 0.39 & \scriptsize\textcolor{gray!60}{$\pm$0.01} \\
 & Scroll Btm & 0.33 & \scriptsize\textcolor{gray!60}{$\pm$0.00} & 0.33 & \scriptsize\textcolor{gray!60}{$\pm$0.00} \\
\bottomrule
\end{tabular}

}
  \end{subtable}
  \hfill
  \begin{subtable}{0.49\textwidth}
    \centering
\scalebox{0.8}{%
  \begin{tabular}{llrlrl}
\toprule
& & \multicolumn{4}{c}{Wall-clock time (s)} \\
\cmidrule(lr){3-6}
 &  & \multicolumn{2}{c}{Before elision} & \multicolumn{2}{c}{After elision} \\
\midrule
\multirow{4}{*}{\rotatebox{90}{\grmtools}} & Eclipse & 5.09 & \scriptsize\textcolor{gray!60}{$\pm$0.05} & 3.59 & \scriptsize\textcolor{gray!60}{$\pm$0.06} \\
 & Hadoop & 4.49 & \scriptsize\textcolor{gray!60}{$\pm$0.03} & 3.07 & \scriptsize\textcolor{gray!60}{$\pm$0.05} \\
 & Jenkins & 4.45 & \scriptsize\textcolor{gray!60}{$\pm$0.02} & 3.02 & \scriptsize\textcolor{gray!60}{$\pm$0.01} \\
 & Spring & 4.11 & \scriptsize\textcolor{gray!60}{$\pm$0.03} & 2.65 & \scriptsize\textcolor{gray!60}{$\pm$0.01} \\
\midrule
\multirow{13}{*}{\rotatebox{90}{\ripgrep}} & Alternates & 1.44 & \scriptsize\textcolor{gray!60}{$\pm$0.03} & 1.39 & \scriptsize\textcolor{gray!60}{$\pm$0.01} \\
 & Alternates (-i) & 1.55 & \scriptsize\textcolor{gray!60}{$\pm$0.02} & 1.50 & \scriptsize\textcolor{gray!60}{$\pm$0.01} \\
\bottomrule
\end{tabular}

      }
  \end{subtable}
  \label{tab:app:elision:wallclock}
\end{table}

\clearpage

\begin{table}[t]
    \centering
    \caption{User times (seconds) for each benchmark in the \Eelision
    experiment, shown before and after applying \ourgc's finalizer elision
    optimisation. Values show arithmetic means over \benchmarkpexecs runs, with
    99\% confidence intervals.}
  \begin{subtable}{0.49\textwidth}
    \centering
\scalebox{0.8}{%
  \begin{tabular}{llrlrl}
\toprule
& & \multicolumn{4}{c}{User time (s)} \\
\cmidrule(lr){3-6}
 &  & \multicolumn{2}{c}{Before elision} & \multicolumn{2}{c}{After elision} \\
\midrule
\multirow{10}{*}{\rotatebox{90}{\alacritty}} & Cur. Motion & 1.08 & \scriptsize\textcolor{gray!60}{$\pm$0.06} & 1.12 & \scriptsize\textcolor{gray!60}{$\pm$0.05} \\
 & Dense Cells & 6.70 & \scriptsize\textcolor{gray!60}{$\pm$0.12} & 6.68 & \scriptsize\textcolor{gray!60}{$\pm$0.13} \\
 & Light Cells & 0.34 & \scriptsize\textcolor{gray!60}{$\pm$0.04} & 0.33 & \scriptsize\textcolor{gray!60}{$\pm$0.04} \\
 & Scroll & 0.12 & \scriptsize\textcolor{gray!60}{$\pm$0.01} & 0.12 & \scriptsize\textcolor{gray!60}{$\pm$0.01} \\
 & Scroll (fullscreen) & 0.27 & \scriptsize\textcolor{gray!60}{$\pm$0.03} & 0.26 & \scriptsize\textcolor{gray!60}{$\pm$0.03} \\
 & Scroll Btm & 0.15 & \scriptsize\textcolor{gray!60}{$\pm$0.01} & 0.15 & \scriptsize\textcolor{gray!60}{$\pm$0.01} \\
\bottomrule
\end{tabular}

}
  \end{subtable}
  \hfill
  \begin{subtable}{0.49\textwidth}
    \centering
\scalebox{0.8}{%
  \begin{tabular}{llrlrl}
\toprule
& & \multicolumn{4}{c}{User time (s)} \\
\cmidrule(lr){3-6}
 &  & \multicolumn{2}{c}{Before elision} & \multicolumn{2}{c}{After elision} \\
\midrule
\multirow{4}{*}{\rotatebox{90}{\grmtools}} & Eclipse & 5.81 & \scriptsize\textcolor{gray!60}{$\pm$0.07} & 3.42 & \scriptsize\textcolor{gray!60}{$\pm$0.06} \\
 & Hadoop & 5.27 & \scriptsize\textcolor{gray!60}{$\pm$0.04} & 2.96 & \scriptsize\textcolor{gray!60}{$\pm$0.05} \\
 & Jenkins & 5.21 & \scriptsize\textcolor{gray!60}{$\pm$0.04} & 2.86 & \scriptsize\textcolor{gray!60}{$\pm$0.01} \\
 & Spring & 4.94 & \scriptsize\textcolor{gray!60}{$\pm$0.05} & 2.54 & \scriptsize\textcolor{gray!60}{$\pm$0.01} \\
\midrule
\multirow{13}{*}{\rotatebox{90}{\ripgrep}} & Alternates & 0.58 & \scriptsize\textcolor{gray!60}{$\pm$0.02} & 0.50 & \scriptsize\textcolor{gray!60}{$\pm$0.02} \\
 & Alternates (-i) & 0.72 & \scriptsize\textcolor{gray!60}{$\pm$0.02} & 0.64 & \scriptsize\textcolor{gray!60}{$\pm$0.02} \\
\bottomrule
\end{tabular}

      }
  \end{subtable}
    \label{tab:app:elision:user}
\end{table}

\begin{table}[t]
    \centering
    \caption{Average heap footprint (MiB) across benchmarks in the \Eelision
    experiment, measured before and after applying \ourgc’s finalizer elision
    optimisation. Values are arithmetic means over \benchmarkpexecs runs with
    resampled traces from heaptrack, with 99\% confidence intervals (see
    \cref{sec:datapres} for details).}
    \label{tab:app:elision:heap}
  \begin{subtable}{0.47\textwidth}
    \centering
\scalebox{0.81}{%
  \begin{tabular}{llrlrl}
\toprule
& & \multicolumn{4}{c}{Avg. heap footprint (MiB)} \\
\cmidrule(lr){3-6}
 &  & \multicolumn{2}{c}{Before elision} & \multicolumn{2}{c}{After elision} \\
\midrule
\multirow{10}{*}{\rotatebox{90}{\alacritty}} & Cur. Motion & 3.76 & \scriptsize\textcolor{gray!60}{$\pm$0.06} & 3.79 & \scriptsize\textcolor{gray!60}{$\pm$0.06} \\
 & Dense Cells & 3.74 & \scriptsize\textcolor{gray!60}{$\pm$0.06} & 3.74 & \scriptsize\textcolor{gray!60}{$\pm$0.06} \\
 & Light Cells & 3.76 & \scriptsize\textcolor{gray!60}{$\pm$0.08} & 3.73 & \scriptsize\textcolor{gray!60}{$\pm$0.07} \\
 & Scroll & 3.88 & \scriptsize\textcolor{gray!60}{$\pm$0.13} & 3.87 & \scriptsize\textcolor{gray!60}{$\pm$0.12} \\
 & Scroll (fullscreen) & 6.46 & \scriptsize\textcolor{gray!60}{$\pm$0.28} & 6.46 & \scriptsize\textcolor{gray!60}{$\pm$0.34} \\
 & Scroll Btm & 3.72 & \scriptsize\textcolor{gray!60}{$\pm$0.07} & 3.73 & \scriptsize\textcolor{gray!60}{$\pm$0.07} \\
\bottomrule
\end{tabular}

}
  \end{subtable}
  \hfill
  \begin{subtable}{0.47\textwidth}
    \centering
\scalebox{0.79}{%
  \begin{tabular}{llrlrl}
\toprule
& & \multicolumn{4}{c}{Avg. heap footprint (MiB)} \\
\cmidrule(lr){3-6}
 &  & \multicolumn{2}{c}{Before elision} & \multicolumn{2}{c}{After elision} \\
\midrule
\multirow{4}{*}{\rotatebox{90}{\grmtools}} & Eclipse & 1993.86 & \scriptsize\textcolor{gray!60}{$\pm$9.04} & 1069.95 & \scriptsize\textcolor{gray!60}{$\pm$28.40} \\
 & Hadoop & 1882.88 & \scriptsize\textcolor{gray!60}{$\pm$1.55} & 800.15 & \scriptsize\textcolor{gray!60}{$\pm$48.62} \\
 & Jenkins & 1876.35 & \scriptsize\textcolor{gray!60}{$\pm$14.92} & 962.16 & \scriptsize\textcolor{gray!60}{$\pm$29.20} \\
 & Spring & 1752.72 & \scriptsize\textcolor{gray!60}{$\pm$1.68} & 1036.27 & \scriptsize\textcolor{gray!60}{$\pm$19.48} \\
\midrule
\multirow{13}{*}{\rotatebox{90}{\ripgrep}} & Alternates & 28.23 & \scriptsize\textcolor{gray!60}{$\pm$1.59} & 15.80 & \scriptsize\textcolor{gray!60}{$\pm$0.40} \\
 & Alternates (-i) & 29.52 & \scriptsize\textcolor{gray!60}{$\pm$1.66} & 17.50 & \scriptsize\textcolor{gray!60}{$\pm$0.66} \\
\bottomrule
\end{tabular}

      }
  \end{subtable}
\end{table}

\end{document}